\documentclass[journal,twocolumns]{IEEEtran}

\usepackage{graphicx}
\DeclareGraphicsExtensions{.png,.eps,.ps,.pdf,.jpg}
\usepackage{url}
\usepackage{color}
\usepackage{cite}
\usepackage{acronym}
\usepackage{epstopdf}
\usepackage{bm}
\usepackage{booktabs}
\usepackage{amsmath}
\usepackage{amssymb}
\usepackage{mathabx}
\usepackage{csquotes}
\usepackage{algorithm}
\usepackage{siunitx}
\usepackage[noend]{algpseudocode}
\usepackage{algpseudocode,algorithm,algorithmicx}

\usepackage{caption}
\usepackage{cespvect}
\usepackage{import}
\usepackage{multirow}
\usepackage{cite}
\usepackage[export]{adjustbox}
\usepackage{breqn}
\usepackage{mathrsfs}
\usepackage{acronym}
\usepackage[ansinew]{inputenc} 
\usepackage{algpseudocode}
\usepackage{algpseudocode,algorithm,algorithmicx}

\usepackage{caption}
\usepackage{subcaption}
\renewcommand{\thetable}{\arabic{table}}

\acrodef{WPT}[WPT]{Wireless Power Transfer}
\acrodef{BS}[BS]{Base Station}
\acrodef{PPG}[PPG]{Power Packet Grid}
\acrodef{DC}[DC]{Direct Current}
\acrodef{TDM}[TDM]{Time Division Multiplexing}
\acrodef{MPC}[MPC]{Model Predictive Control}
\acrodef{GP}[GP]{Gaussian Process}
\acrodef{GPs}[GPs]{Gaussian Processes}
\acrodef{CO}[CO]{Convex Optimization}
\acrodef{TIM}[TIM]{Telecom Italia Mobile}
\acrodef{EB}[EB]{Energy Buffer}
\acrodef{QoS}[QoS]{Quality of Service}
\acrodef{EH}[EH]{Energy Harvesting}

\definecolor{ao}{rgb}{0.0, 0.5, 0.0}
\definecolor{antiquefuchsia}{rgb}{0.57, 0.36, 0.51}

\newcommand{\nonparam}{\mbox{non-parametric}}
\newcommand{\realworld}{\mbox{real-world}}
\newcommand{\nonlinear}{\mbox{non-linear}}
\newcommand{\timeseries}{\mbox{time-series}}

\newcommand{\po}{\gamma}

\newcommand{\bmax}{B_{\max}}
\newcommand{\bup}{B_{\rm up}}
\newcommand{\bref}{B_{\rm ref}}
\newcommand{\blow}{B_{\rm low}}
\newcommand{\seton}{\mathcal{S}_{\rm on}}
\newcommand{\setoff}{\mathcal{S}_{\rm off}}
\newcommand{\eq}[1]{Eq.~(\ref{#1})}
\newcommand{\eqsimple}[1]{(\ref{#1})}
\newcommand{\secref}[1]{Section~\ref{#1}}

\newcommand{\myhat}[1]{\widehat{#1}}
\newcommand{\mybm}[1]{\bm{#1}}

\usepackage{mdframed}
\usepackage{tikz}

\usepackage[keeplastbox]{flushend}

\graphicspath{{./figure/}}
\begin{document}

\title{Energy Cooperation for Sustainable Base Station Deployments: Principles and Algorithms}
\title{Energy Sustainable Mobile Networks via Energy Routing, Learning and Foresighted Optimization}

\author{\IEEEauthorblockN{\'Angel Fern\'andez Gamb\'in, Maria Scalabrin, Michele Rossi}}

\maketitle

\begin{abstract}
The design of \mbox{self-sustainable} base station (BS) deployments is addressed in this paper: BSs have energy harvesting and storage capabilities, they can use ambient energy to serve the local traffic or store it for later use. A dedicated power packet grid allows energy transfer across BSs, compensating for imbalance in the harvested energy or in the traffic load. Some BSs are {\it offgrid}, i.e., they can only use the locally harvested energy and that transferred from other BSs, whereas others are {\it ongrid}, i.e., they can also purchase energy from the power grid. Within this setup, an optimization problem is formulated where: energy harvested and traffic processes are estimated at the BSs through Gaussian Processes (GPs), and a Model Predictive Control (MPC) framework is devised for the computation of energy allocation and transfer schedules. Numerical results, obtained using real energy harvesting and traffic profiles, show substantial improvements in terms of energy \mbox{self-sustainability} of the system, outage probability (zero in most cases), and in the amount of energy purchased from the power grid, which is of more than halved with respect to the case where the optimization does not consider GP forecasting and MPC.
\end{abstract}



\section{Introduction}
\label{sec:intro}

The massive use of Information and Communications Technologies (ICT) is increasing the amount of energy drained by the telecommunication infrastructure and its footprint on the environment. Forecast values for 2030 are that $51\%$ of the global electricity consumption and $23\%$ of the carbon footprint by human activity will be due to ICT~\cite{andrae2015global}. As such, energy efficiency and \mbox{self-sufficiency} are becoming key considerations for any development in the ICT sector.

In this paper, we advocate future networks where small Base Stations (BSs) are densely deployed to offer coverage and high data rates, and energy harvesting hardware (e.g., solar panels and energy storage units) is also installed to power them~\cite{zordan2015telecommunications}. These BSs collect energy from the environment, use it to serve their local traffic and {\it transfer} it to other BSs to compensate for imbalance in the harvested energy or in the traffic load. Some of the \acp{BS} are connected to the power grid (referred to as {\it ongrid}), whereas the others are {\it offgrid} and, as such, rely on either the locally harvested energy or on the energy transferred from other \acp{BS}. Since the \acp{BS} have a local energy storage, they can accumulate energy when the harvested inflow is abundant. Moreover, some of the surplus energy can be transferred to other \acp{BS} to ensure the \mbox{self-sustainability} of the cellular system. Energy transfer is a prime feature of these networks and can be accomplished in two ways: i) through \ac{WPT} or ii) using a \ac{PPG}~\cite{sugiyama2012packet}. For i), previous studies~\cite{Bonati-2017} have shown that its transfer efficiency is too low for it to be a viable solution when distances exceed a few meters, but ii) looks promising. In analogy with communications networks, in a \ac{PPG} a number of power {\it sources} and power {\it consumers} exchange power (Direct Current, DC) in the form of ``packets'', which flow from sources to consumers thanks to power lines and electronic switches. The energy routing process is controlled by a special entity called the \textit{energy router}~\cite{krylov2010toward}. Following this architecture, a local area packetized power network consisting of a group of energy subscribers and a core energy router is presented in~\cite{ma2017optimal}, where a strategy to match energy suppliers and consumers is devised.


Within this setting, in this paper the allocation and distribution of energy is performed through the \ac{PPG} infrastructure, where a centralized energy router is responsible for deciding the power allocation/transfer among \acp{BS} over time, referred to here as {\it energy routing}. This energy allocation and transfer problem is solved combining \ac{GPs}, \ac{MPC} and \ac{CO}. \ac{GPs} are utilized to learn the \ac{EH} and consumption patterns, which are then fed into \ac{MPC} and \ac{CO} techniques to obtain energy distribution schedules across subsequent time slots, solving a finite horizon optimization problem. This framework is designed for online use and combines learning and foresighted optimization. Numerical results, obtained with \realworld\ harvested energy traces and traffic load patterns, show that the proposed approach effectively keeps the outage probability\footnote{Computed as the ratio between the number of \acp{BS} that are unable to serve the users within range due to energy scarcity, and the total number of \acp{BS}.} to nearly zero for a wide range of traffic loads and system configurations. Also, the amount of energy purchased from the power grid to operate the mobile network is reduced of more than $50$\% with respect to the case where energy schedules are computed solely based on the current network status, i.e., disregarding future energy arrivals and load conditions. The proposed approach extends our previous work in~\cite{Gambin2017}, adding online learning features and foresighted optimization (via \ac{MPC}), whose combination is here proven to lead to substantial improvements.

The paper is organized as follows. In Section~\ref{sec:related_work}, the literature on energy cooperation and the mathematical tools used in this work are presented. Section~\ref{sec:systemModel} describes the network scenario. The overall optimization framework for online energy management is explained in Section~\ref{sec:problem_form}, where the proposed solutions are also detailed. Routing and scheduling policies are addressed in Section~\ref{sec:scheduling}. The numerical results are presented in section~\ref{sec:results}, whereas final remarks are given in section~\ref{sec:conclusions}.



\section{Related Work}
\label{sec:related_work}

In this section, we first survey the main literature dealing with energy transfer in mobile networks, and then delve into the description of the mathematical tools that we consider in this paper, discussing their successful use within diverse application domains. \\

\noindent \textbf{Energy transfer in mobile cellular networks:} the concept of {\it energy transfer}, also referred to as energy cooperation~\cite{chia2014energy, Gurakan2013, Xu2015} or energy exchange~\cite{leithon2014energy}, is motivated by the fact that the distributed renewable energy generated at the base stations can be leveraged upon through a microgrid connecting the BSs~\cite{farooq2017hybrid}, with the aim of improving the network \mbox{self-sustainability}, while reducing the cost entailed in purchasing the energy from the main power grid. Since this is a rather new paradigm, only few works dealing with this problem have been published so far.
Energy sharing among \acp{BS} is investigated in~\cite{Gurakan2013} through the analysis of several multiuser network structures. A \mbox{two-dimensional} and directional \mbox{water-filling-based} and offline algorithm is proposed to control the harvested energy flows in time and space, with the objective of maximizing the system throughput. In~\cite{Xu2015}, the authors introduce a new entity called the \textit{aggregator}, which mediates between the grid operator and a group of \acp{BS} to redistribute the energy flows, reusing the existing power grid infrastructure: one \ac{BS} injects power into the aggregator and, simultaneously, another one draws power from it. This scheme does not consider the presence of energy storage devices, and for this reason some of the harvested energy can be lost if none of the base stations needs it at a certain time instant. The proposed algorithm tries to jointly optimize the transmit power allocations and the transferred energy, maximizing the \mbox{sum-rate} throughput for all the users.
The authors of~\cite{huang2017smart} consider BSs with energy harvesting capabilities connected to the power grid as a means to carry out the energy trading. A joint optimization tackling BS operation and power distribution is performed to minimize the \mbox{on-grid} power consumption of the BSs. Wired energy transfer to/from the power distribution network, and a \mbox{user-BS} association scheme based on cell zooming are investigated. The problem is split into two subproblems, which are solved using heuristics.
A similar approach is considered in~\cite{han2013optimizing}, where two problems are solved: the first one consists of optimizing the  energy allocation at individual BSs to accommodate for the temporal dynamics of harvested energy and mobile traffic. Considering the spatial diversity of mobile traffic patterns, the second problem is to balance the energy consumption among BSs, by adapting the cell size (radio coverage) to reduce the \mbox{on-grid} energy consumption of the cellular network. Again, the solutions are obtained through heuristic algorithms. Also, base stations cooperate toward the reduction of energy costs, but do not perform any actual energy transfer among them.

A \mbox{two-cell} \mbox{renewable-energy-powered} system is studied in~\cite{guo2013base}, by maximizing the sum-rate over all users while determining the direction and amount of energy to be transferred between the two BSs. Energy can be transferred across the network either through power lines or wireless transfer and the energy transfer efficiency is taken into account. This resource allocation problem is formulated under a Frequency Division Multiple Access (FDMA) setup and is solved numerically. A \mbox{low-complexity} heuristic approach is also proposed as a practical \mbox{near-optimal} strategy when the transfer efficiency is sufficiently high and the channel gains are similar for all users.

Along the same lines, a \mbox{two-BS} scenario is considered in~\cite{chia2014energy}, where BSs have hybrid conventional (power grid) and renewable energy sources, limited energy storage capability, and are connected through power lines. The authors study the case where renewable energy and energy demand profiles are deterministic or known ahead of time, and find the optimal energy cooperation policy by solving a linear program. They then consider a more realistic case where the profiles are stochastic and propose an online greedy algorithm. Finally, an intermediate scenario is addressed, where the energy profiles are obtained from a deterministic pattern adding a small amount of random noise at each time step (to model prediction errors). Simulation results are shown for several (online) algorithms, assessing the impact of knowing the energy pattern profiles in advance.

The authors of~\cite{farooq2017hybrid} and~\cite{farooq2016energy} consider an energy sharing framework for cellular networks that are powered by power grids and possess renewable energy generation capabilities. Energy sharing takes place via physical power lines, as well as through the power grid for virtual energy transportation. Interestingly, the authors investigate the impact of the power line infrastructure topology: agglomerative and divisive hierarchical clustering algorithms are utilized to determine it. Upon establishing the physical connections among BSs, an optimization framework for \mbox{day-to-day} cost optimization is developed for the cases of 1) zero knowledge, 2) perfect knowledge, and 3) partial future knowledge of the renewable energy generation. 
An optimal energy management strategy to minimize the energy cost incurred by a set of cellular base stations is presented in~\cite{leithon2014energy}. There, base stations can exchange energy with the power grid and are equipped with batteries (energy storage) and renewable energy harvesting devices. Simulation results show that a cost reduction can be achieved by increasing the battery capacity of each BS and/or the number of base stations.\\

\noindent \textbf{On combining pattern learning with \mbox{multi-step} optimization techniques:} next, we briefly review the mathematical tools that we use in the present paper, namely, \ac{MPC} and \ac{GPs}, touching upon the various application domains where they have been used.
%
%
\ac{MPC} has its roots in optimal control theory. The main idea is to use a dynamic model to forecast the system behavior, and exploit the forecast state sequence to obtain the {\it control} at the current time. The system usually evolves in slotted time, the control action is obtained by solving, at each time step, a finite horizon optimal control problem where the initial state is the current state of the system. The optimization yields a finite control sequence, and the first control action in this sequence is applied~\cite{rawlings2009model}. \ac{MPC} has the ability to anticipate future events and can take control actions accordingly. It has been widely used in industrial processes, including chemical plants~\cite{fang2015nonlinear, eaton1991model, arefi2006nonlinear} and oil refineries~\cite{nejadkazemi2016pressure, pavlov2014modelling} and, recently, to balance energy consumption in smart energy grids~\cite{halvgaard2016distributed, stadler2016distributed, meng2015cooperation}. Moreover, it has been applied to supply chain management problems, with promising results~\cite{tzafestas1992parallel, perea2003model, braun2003model, lin2005predictive}. 

It is known that using \mbox{time-series} forecasting within an \ac{MPC} framework can improve the quality of the control actions by providing insight into the future~\cite{doganis2008combined}. Over the last decades, numerous forecasting approaches have been developed, including Autoregressive Integrated Moving Average (ARIMA) processes and Artificial Neural Networks (ANNs). ARIMA models (introduced by Box and Jenkins in~\cite{box1970}) are known for their prediction accuracy, but their main limitation lies in the assumption that the data follows a linear model. Conversely, ANNs capture \nonlinear\ models and, in turn, can be a good alternative to ARIMA~\cite{zhang2005neural}. Nonetheless, ANNs give rise to mixed results for purely linear correlation structures. In~\cite{zhang2003time,khandelwal2015time}, hybrid schemes that combine them are put forward to take advantage of their unique strengths. Experimental results with \mbox{real-world} data indicate that their combined use can improve the prediction accuracy achieved by either of the techniques when used in isolation. 

Several authors have proposed the use of \nonlinear\ models to build \nonlinear\ adaptive controllers. In most applications, however, these non-linearities are unknown, and \nonlinear\ parameterization must be used instead. In \timeseries\ analysis, where the underlying structure is largely unknown, one of the main challenges is to define an appropriate form of \nonlinear\ parameterization for the forecasting model. Some implementations claim to be \nonparam, such as \acp{GPs}, which can be considered (in some sense) as equivalent to models based on an infinite set of \nonlinear\ basis functions~\cite{mackay1997gaussian}. The basic idea of \ac{GPs} is to place a \emph{prior distribution} directly on the space of functions, without finding an appropriate form of \nonlinear\ parameterization for the forecasting model. This can be thought of as a generalization of a Gaussian distribution over functions. Moreover, a \ac{GP} is completely specified by the mean function and by the \emph{covariance function} or \emph{kernel}, which has a particular (but simple) parametric structure, defined through a small number of \emph{hyperparameters}. The term \nonparam\ does not mean that there are no parameters, but that the parameters can be conveniently adapted from data. While \ac{GPs} have been used in \timeseries\ forecasting~\cite{williams1998prediction}, to the best of the authors' knowledge, \cite{leith2004gaussian} is the first application of \ac{GPs} to electrical load forecasting~\cite{blum2013electricity, taylor2003short, taylor2006comparison, ketter2013power}. 
The electricity demand is mainly influenced by meteorological conditions and daily seasonality. Nevertheless, forecasting for \mbox{short-term} horizons of about a day is often performed using univariate prediction models, which are considered to be sufficient because the weather tends to change in a smooth fashion, which is reflected in the electricity demand itself. Also, in a \realworld online forecasting scenario, multivariate modeling is usually considered impractical~\cite{taylor2003}. Due to daily seasonality, we can say that the electrical load data bears some similarities with the time series that we consider in this paper, i.e., the harvested energy profile of \secref{sub:harvested_energy} and the traffic load of \secref{sub:traffic_load}.   

The idea of combining \ac{MPC} and \ac{GP} was first proposed in~\cite{kocijan2003predictive}. Other practical implementations can be found in application domains such as greenhouse temperature control systems~\cite{pawlowski2011predictive}, \mbox{gas-liquid} separation plant control systems~\cite{likar2007predictive}, combustion power plants control systems~\cite{grancharova2008explicit} and in a number of other cases~\cite{palm2007multiple,ko2007gaussian,maciejowski2013fault,wang2014gaussian}. To the best of our knowledge, the present work is the first where \ac{MPC} and \ac{GP} are combined to control an energy harvesting mobile network. 


\section{System Model}
\label{sec:systemModel}

We consider a mobile network comprising a set $\mathcal{S}$ of $n_s = |\mathcal{S}|$ \acp{BS}, each with energy harvesting capabilities, i.e., a solar panel, an energy conversion module and an energy storage device. Some of the \acp{BS} are ongrid (termed {\it ongrid} \acp{BS}, being part of set $\mathcal{S}_{\rm on}$) and, in turn, can obtain energy from the power grid. The remaining \acp{BS} are {\it offgrid} (set $\mathcal{S}_{\rm off}$). The proposed optimization process evolves in slotted time $t=1,2,\dots$, where the slot duration corresponds to the time granularity of the control and can be changed without requiring any modifications to the following algorithms.

\subsection{Power Packet Grids}
\label{electrical_grid}

\begin{figure}[t]
	\centering
	\includegraphics[width=\columnwidth]{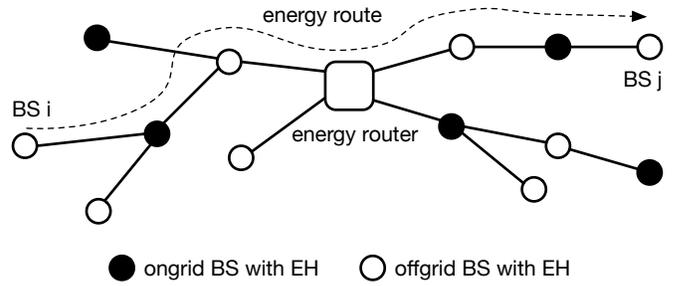}
	\caption{Power packet grid topology example.}
	\label{fig:PPG}
\end{figure}

A \ac{PPG} is utilized to distribute energy among the \acp{BS}. The grid architecture is similar to that of a \mbox{multi-hop} network, see Fig.~\ref{fig:PPG}, where circles are \acp{BS} and the square is the energy router, which is in charge of energy routing decisions and power allocation. As assumed in~\cite{ma2017optimal}, \acp{BS} are connected through \ac{DC} power links (electric wires) and the transmission of energy over them is operated in a \ac{TDM} fashion. Energy transfer occurs by first establishing an {\it energy route}, which corresponds to a sequence of power links between the energy source and the destination. Each power link can only be used for a single transfer operation at a time. Power distribution losses along the power links follow a linear function of the distance between the source and the destination~\cite{ma2017optimal}. They depend on the resistance of the considered transmission medium and are defined by~\cite{von2006electric}: $R = \rho \ell / A$, where $\rho$ is the resistivity of the wire in $\SI{}{\Omega \milli\milli^2 / \meter}$, $\ell$ is the length of the power link in meters, and $A$ is the cross-sectional area of the cable in $\SI{}{\milli\milli^2}$. In this paper, we consider a \ac{PPG} with a single energy router in the center of the topology. A number of trees originates from the router and, without loss of generality, each hop is assumed to have the same length $\ell$, i.e., the same power loss.

\subsection{Harvested Energy Profiles}
\label{sub:harvested_energy}

Solar energy generation traces have been obtained using the SolarStat tool~\cite{miozzo2014solarstat}. For the solar modules, the commercially available Panasonic N235B photovoltaic technology is considered. Each solar panel has $25$ solar cells, leading to a panel area of $\SI{0.44}{\meter}^2$, which is deemed practical for installation in a urban environment, e.g., on light-poles. As discussed in~\cite{miozzo2014solarstat,zordan2015telecommunications}, the \ac{EH} inflow is generally \mbox{bell-shaped} with a peak around \mbox{mid-day}, whereas the energy harvested during the night is negligible. 
Here, the framework in~\cite{miozzo2014solarstat} is utilized to obtain the amount of energy harvested for each \ac{BS}  \mbox{$n = 1, \dots, n_s$} in time slot $t$, which is denoted by $H_n(t)$.

\subsection{Traffic Load and Power Consumption}
\label{sub:traffic_load}

Traffic load traces have been obtained using real mobile data from the Big Data Challenge organized by \ac{TIM}~\cite{bigdata2015tim}. The dataset is the result of a computation over Call Detail Records (CDRs), logging the user activity within the TIM cellular network for the city of Milan during the months of November and December 2013. For the traffic load traces we use the CDRs related to SMS, calls and Internet activities, performing spatial and temporal aggregation. In this way, we obtain a daily traffic load profile for each BS.

\begin{figure}[t]	
		\centering
		\includegraphics[width=\columnwidth]{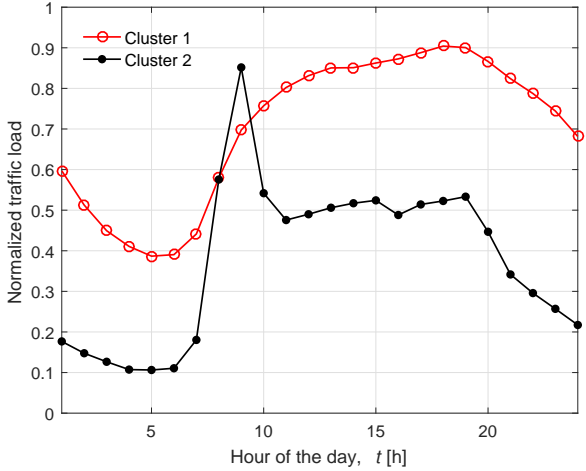}
		\caption{Load pattern profiles (two classes).}
		\label{fig:clusters}	
\end{figure}

Clustering techniques have been applied to the dataset to understand the behavior of the mobile data. To this end, we use DBSCAN unsupervised clustering~\cite{ester1996density} to classify the load profiles into several categories. In Fig.~\ref{fig:clusters}, we show the typical traffic behavior of two clusters, corresponding to the heaviest (cluster 1) and lightest (cluster 2) daily load. As noted in previous works, the traffic is \mbox{time-correlated} (and daily periodic)~\cite{zordan2015telecommunications, auer2010d2}. 
In our numerical results, each \ac{BS} has an associated load profile, which is picked at random as one of the two clusters in Fig.~\ref{fig:clusters}. Depending on the cluster association probabilities, there is some imbalance in the load distribution across \acp{BS} that, as we discuss shortly, plays a key role in the performance of energy transfer algorithms. Given the traffic load profile $L_n(t)$, intended as the percentage of the total bandwidth that a \ac{BS} $n$ allocates to serve the users in its radio cell, the \ac{BS} energy consumption (energy outflow), referred to in the following as $O_n(t)$, is computed through the linear model in~\cite{zordan2015telecommunications} (see Eq.~(1) in that paper).

\subsection{Energy Storage Units}
\label{sub:storage_units}


Energy storage units are interchangeably referred to as \acp{EB}. The \ac{EB} level for BS $n=1,\dots,n_s$ is denoted by $B_n(t)$ and three thresholds are defined: $\bup$, $\bref$ and $\blow$, respectively termed the {\it upper}, {\it reference} and {\it lower} energy threshold, with $0 < \blow < \bref < \bup < \bmax$ ($\bmax$ is the \ac{EB} capacity). $\bref$ corresponds to the desired \ac{EB} level and $\blow$ is the lowest energy level that any BS should ever reach. Both variables are used in the optimization of Section~\ref{sub:model_predictive_control}. For an offgrid \ac{BS}, i.e., $n \in \setoff$, if $t$ is the current time slot, the buffer level process is updated at the beginning of the next time slot $t+1$ as:
\begin{equation}
	\label{eq:buffer_1}
	B_n(t+1) = B_n(t) + H_n(t) - O_n(t) + T_n(t) \, ,
\end{equation}
where $T_n(t)$ is the amount of energy transferred to/from \ac{BS} $n$ in time slot $t$, which is positive if \ac{BS} $n$ is a consumer or negative if \ac{BS} $n$ acts as an energy source. $B_n(t)$ is the \ac{EB} level at the beginning of time slot $t$, whereas $H_n(t)$, $O_n(t)$ are the amount of energy harvested and the energy that is locally drained (to support the local data traffic), respectively.
%
The energy level of an ongrid BS $n \in \seton$ is updated as: 
\begin{equation}
	\label{eq:buffer_2}
	B_n(t+1) = B_n(t) + H_n(t) - O_n(t) + T_n(t) + \theta_n(t) \, ,
\end{equation}
where $\theta_n(t) \geq 0$ represents the energy purchased by \ac{BS} $n$ from the power grid during time slot $t$. The behavior of a \ac{BS} (i.e., $T_n(t)$ and $\theta_n(t)$) depends on its \ac{EB} level. If the BS behaves as an {\it energy source}, it is eligible for transferring a certain amount of energy $T_n(t)$ to other \acp{BS}. In this work, we assume that if the total energy in the buffer at the beginning of the current time slot $t$ is $B_n(t) < \bup$ and the BS $n$ is ongrid, then the difference $\theta_n(t) = \bup - B_n(t)$ is purchased from the power grid in slot $t$, as an ongrid BS should always be a source, i.e., in the position of transferring energy to other \acp{BS}. If instead the BS behaves as an {\it energy consumer}, it demands energy to the sources. For example, the energy demand in time slot $t$ may be set to \mbox{$\bref - B_{n}(t)$}, so that the \ac{EB} level would ideally become no smaller than the reference threshold $\bref$ by the end of the current time slot $t$. Note that, this can only be strictly guaranteed if $H_n(t) - O_n(t) \geq 0$. However, $B_n(t)$ is updated at the beginning of time slot $t$, whereas $H_n(t)$ and $O_n(t)$ are only known at the end of it. The theory of Sections~\ref{sub:prediction} and~\ref{sub:model_predictive_control} allows computing $T_n(t)$, accounting for the expected behavior \mbox{$\mathbb{E}[H_n(t) - O_n(t)]$} to get more accurate results, where $\mathbb{E}[\cdot]$ is the expectation operator.



\begin{figure*}[t]
	\centering
	\includegraphics[width=0.7\textwidth]{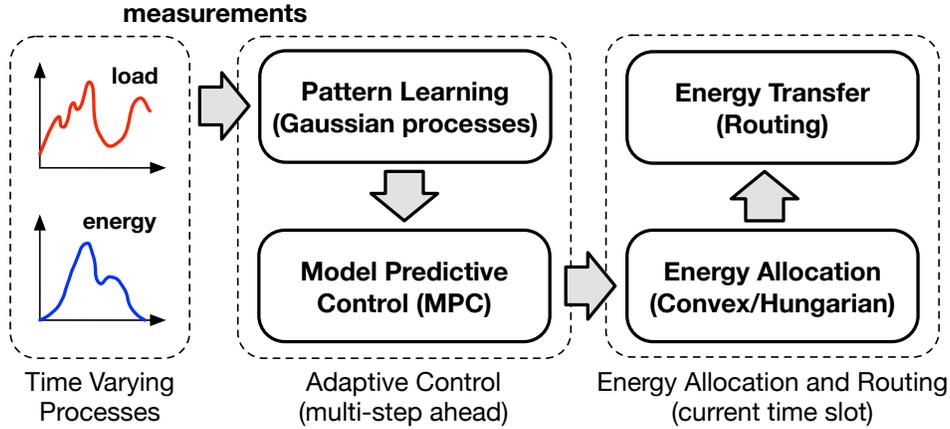}
	\caption{Overview of the optimization framework.}
	\label{fig:opt_framework}
\end{figure*}

\section{Optimization for online energy management}
\label{sec:problem_form}

In this section, we devise an online optimal power allocation strategy, whose objective is to make the offgrid \acp{BS} as energy \mbox{self-sustainable} as possible. This is achieved by transferring some amount of energy from rich energy \acp{BS} (energy sources) to those base stations that require energy (energy consumers). 

\subsection{Overview of the optimization framework}
\label{sub:framework}

A diagram of the optimization process is shown in Fig.~\ref{fig:opt_framework}, involving 1) pattern learning (forecasting), 2) model predictive control, 3) convex optimization and 4) energy routing. These algorithms are all executed at runtime. First of all, the harvested energy and traffic load processes are statistically modeled through Bayesian \nonparam\ tools (``pattern learning'' in Fig.~\ref{fig:opt_framework}), as we detail in Section~\ref{sub:prediction}. This first step allows each \ac{BS} to independently track its own energy and load processes, capturing their statistical behavior and obtaining \mbox{multi-step} ahead forecasts for the corresponding time series. It is worth noting that our forecasting method is agnostic to the type of signal, and for this reason can be promptly extended to other processes, if need be. These forecasts are then fed into the foresighted optimization approach of Section~\ref{sub:model_predictive_control}. Their use allows for informed decisions, which take the future system evolution into account. This results in effective energy allocation strategies, which lead to a reduction of the amount of energy that has to be purchased from the power grid.

The second step in the optimization framework is the adaptive control based on \ac{MPC} theory. Its main goal is to determine the \ac{BS} role (energy source or consumer), and obtain the amount of energy $T_n(t)$ that each \ac{BS} $n$ has to either transfer or  require from the sources. The \ac{MPC} block takes online actions, considering not only the current system state, i.e., traffic load, harvested energy and \ac{EB} levels, but also future ones (based on the forecasts from the previous block), anticipating events and acting accordingly. This is a main difference with respect to the work in~\cite{Gambin2017}, where \ac{BS} energy roles are solely determined on the current \ac{EB} level. The \ac{MPC} block is described in Section~\ref{sub:model_predictive_control}.

The actual energy allocation is evaluated in the third optimization step. This block computes how energy $T_n(t)$ (obtained through \ac{MPC}) has to be redistributed among \acp{BS}, matching energy sources and energy consumers. Two approaches are proposed to this end (see Section~\ref{sub:convex_optimization}): one based on convex optimization and another one formulated as an assignment problem and solved through the Hungarian method~\cite{kuhn1955hungarian}. Their objective is to reduce as much as possible the outage probability, i.e., the ratio between the number of \acp{BS} that are unable to serve their load due to energy scarcity, and the total number of \acp{BS} in the network, while maximizing the energy transfer efficiency.

Finally, the last step is to perform the energy exchange (``energy routing'') among the \acp{BS}. Since the PPG is operated in a \ac{TDM} fashion, each power link can only be used for a single trading operation at a time. Hence, the proposed routing strategy seeks to find disjoint routes between energy sources and consumers, while minimizing the time needed to perform the energy transfer. Details are provided in Section~\ref{sec:scheduling}. 

\subsection{Pattern learning through Bayesian \nonparam\ models}
\label{sub:prediction}

\label{sub:GP}

\begin{table}[t]
	\centering
	\caption{List of symbols used in the paper.}
	\label{tbl:sensor}
	\begin{tabular}{ | l | l | }
		\hline
		\textbf{Definition} & \textbf{Variable name} \\   \hline
		Base station set & $\mathcal{S}$ \\
		Ongrid base station set & $\mathcal{S}_{\rm on}$ \\
		Offgrid base station set & $\mathcal{S}_{\rm off}$ \\
		Number of base stations & $n_s = |\mathcal{S}|$ \\  
		Harvested energy profile in slot $t$ & $H(t)$\\
		Traffic load profile in slot $t$ & $L(t)$\\
		BS energy consumption in slot $t$ & $O(t)$\\
		Energy buffer level in slot $t$ & $B(t)$ \\
		Maximum energy buffer capacity & $\bmax$ \\
		Upper, lower and reference buffer thresholds & $\bup$, $\blow$, $\bref$ \\
		Transferred energy in slot $t$ & $T(t)$\\
		Purchased grid energy in slot $t$ & $\theta(t)$\\ [1mm]
		{\textbf{Used by the pattern analysis block}} & \\
		\hline
		Number of observations (training dataset) & $N$ \\
		Number of observations (test set) & $N_*$ \\
		The transpose of vector $\mybm{x}$ & $\mybm{x}^{\top}$ \\
		The weights of the Bayesian linear model 	& $\mybm{w}$ \\ 
		The function value $f(\mybm{x}) = {\phi(\mybm{x})}^{\top} \mybm{w}$ & $f(\mybm{x})$ \\
		The observed real value $r = f(\mybm{x}) + \epsilon$	 & $r$ \\
		$N$-dimensional column vector of targets & $\mybm{r}$ \\
		Map in the feature space & $\phi(\cdot): \mathbb{R}^D \to \mathbb{R}^F$ \\
		Training dataset $\mathcal{D} = \{(\mybm{x}_i , r_i )\}_{i=1}^N$ & $\mathcal{D}$ \\
		$D$-dimensional input column vector & $\mybm{x}$ \\
		$D$-dimensional input column vector (test set) & $\mybm{x}_*$ \\
		$F$-dimensional feature vector	& $\phi(\mybm{x}) = \mybm{\phi}$ \\
		$D \times N$ matrix of inputs & $\mybm{X}$ \\
		$D \times N_*$ matrix of inputs in the test set & $\mybm{X}_*$ \\
		$F \times N$ matrix in the feature space & $\phi(\mybm{X}) = \mybm{\Phi}$ \\
		Gaussian dist. with zero mean and variance $\sigma_n^2$ & $\epsilon \sim \mathcal{N}(0, \sigma_n^2)$ \\
		Covariance matrix of the model parameters $\mybm{w}$ & $\mybm{\Sigma}_w$ \\
		Gaussian process & $\mathcal{GP}(m(\mybm{x}), k(\mybm{x}, \myhat{\mybm{x}}))$ \\
		Gaussian process: mean function & $m(\mybm{x})$ \\
		Gaussian process: \emph{covariance function (kernel)} & $k(\mybm{x}, \myhat{\mybm{x}})$ \\
		Gaussian process: predictive mean vector & $\mybm{\mu}$ \\
		Gaussian process: predictive covariance matrix & $\mybm{\Sigma}$ \\
		$N \times N$ covariance matrix (training dataset) & $\mybm{K}$ \\
		$N \times N$ identity matrix & $\mybm{I}_N$ \\
		Function values (training dataset) & $\mybm{f}$ \\
		Function values (test set) & $\mybm{f}_*$ \\ [1mm]	
		\textbf{Used by the optimization block} & \\ 
		\hline
		Optimization horizon (time steps) & $M$ \\
		System state matrix for MPC & $\bm{Z}_t$ \\
		Control matrix for MPC & $\bm{U}_t$ \\
		System disturbances matrix for MPC & $\bm{W}_t$ \\
		Weight parameter for MPC & $\alpha$ \\
		Set of energy sources & $\mathcal{Y}_s$ \\
		Set of energy consumers & $\mathcal{Y}_c$ \\
		Energy allocation matrix & $\bm{Y}$ \\
		Energy availability matrix & $\mybm{E}$ \\
		Maximum transmission energy capacity & $e_{\max}$ \\
		Energy demand vector & $\mybm{d}$ \\
		Number of hops matrix & $\bm{G}$ \\
		Weight parameter for energy allocation & $\beta$ \\
		Cost matrix for Hungarian method & $\bm{C}$ \\
		\hline
	\end{tabular}
\end{table}

In this section, we are concerned with statistical models to automatically capture the hidden structure of the observations in a training dataset. Bayesian \mbox{non-parametric} models, such as \ac{GPs}, can represent our beliefs about the model parameters via a probability distribution (called the \emph{prior}). Then, Bayesian inference can reshape the \emph{prior} distribution, transforming it into a \emph{posterior} one.
\acp{GP} have become popular for regression and classification, often showing impressive empirical performance~\cite{Rasmussen_2006}. However, while the outputs for a classification task are discrete class labels, in a regression task the outputs (or targets) are real values. Here, we use \acp{GP} for our regression task. According to~\cite{Rasmussen_2006}, there are two equivalent views to treat \acp{GP} within a regression problem: 1) the \mbox{weight-space} view and 2) the \mbox{function-space} view.\\

\noindent \textbf{1) The weight-space view.} The Bayesian linear model for regression is defined as:
\begin{equation}
\label{eq:1.1}
f(\mybm{x}) = {\phi(\mybm{x})}^{\top} \mybm{w}, \quad r = f(\mybm{x}) + \epsilon,
\end{equation}
where $\mybm{w}$ is a vector of weights, also known as model parameters, $f(\mybm{x})$ is the function value, which is linear in the weights $\mybm{w}$, $r$ is the observed real value, and $\phi(\cdot): \mathbb{R}^D \to \mathbb{R}^F$ maps the \mbox{$D$-dimensional} input column vector $\mybm{x}$ into an \mbox{$F$-dimensional} feature vector $\phi(\mybm{x}) = \boldsymbol{\phi}$. Assume we are given with a training dataset with $N$ observations, $\mathcal{D} = \{(\mybm{x}_i , r_i )\}_{i=1}^N$, where each pair $(\mybm{x}_i , r_i)$ consists of the $D$-dimensional input column vector $\mybm{x}_i$ and the scalar target $r_i$. We can aggregate inputs and targets in a $D \times N$ matrix $\mybm{X}$ and an $N$-dimensional column vector $\mybm{r}$, so that $\mathcal{D} = (\mybm{X}, \mybm{r})$, and $\phi(\mybm{X}) = \boldsymbol{\Phi}$ becomes an $F \times N$ matrix in the feature space. We are interested in the conditional distribution of the targets, given the inputs in the feature space and the model parameters. 
We further assume that $r$ differs from $f(\mybm{x})$ by additive noise, and this noise follows an independent identically distributed (i.i.d.) Gaussian distribution with zero mean and variance $\sigma_n^2$, i.e., \mbox{$\epsilon \sim \mathcal{N}(0, \sigma_n^2)$}. From the i.i.d. assumption, it follows that the \emph{likelihood} (i.e., the conditional distribution of the targets given the inputs in the feature space and the model parameters) is factorized over cases for the $N$ observations, i.e., \mbox{$\mybm{r} | \mybm{X}, \mybm{w} \sim \mathcal{N}({\boldsymbol{\Phi}}^{\top} \mybm{w}, \sigma_n^2\mybm{I}_N)$}.

According to the standard formalism of Bayesian inference, the \emph{prior} distribution encodes our beliefs about the model parameters, before we access the $N$ observations. Conversely, the \emph{likelihood} gives us insights about the model parameters, thanks to the evidence of the observations in a training dataset. 
For the case simple Bayesian linear model of \eq{eq:1.1}, the \emph{prior} distribution is set to be Gaussian with zero mean and covariance matrix $\boldsymbol{\Sigma}_w$, i.e., \mbox{$\mybm{w} \sim \mathcal{N}(\mybm{0}, \boldsymbol{\Sigma}_w)$}. 
Then, the \emph{posterior} distribution combines the \emph{likelihood} and the \emph{prior} distribution, and expresses our full knowledge about the model parameters, after we access the $N$ observations. The \emph{posterior} distribution is derived via the Bayes' rule as:
\begin{equation}
\begin{split}
\label{eq:1.3}
p(\mybm{w} | \mybm{X}, \mybm{r}) = \frac{p(\mybm{r} | \mybm{X}, \mybm{w}) p(\mybm{w})}{p(\mybm{r} | \mybm{X})} =  \frac{p(\mybm{r} | \mybm{X}, \mybm{w}) p(\mybm{w})}{\int p(\mybm{r} | \mybm{X}, \mybm{w}) p(\mybm{w}) d\mybm{w}}.
\end{split}
\end{equation}

To make prediction for the test case $f(\mybm{x}_*) = {f}_*$ given $\phi(\mybm{x}_*) = \boldsymbol{\phi}_*$, we average over all possible model parameters' choices, weighted by the \emph{posterior} distribution, i.e.,
\begin{equation}
\begin{split}
\label{eq:1.4}
& p({f}_* | \mybm{x}_*, \mybm{X}, \mybm{r}) = \int p({f}_* | \mybm{x}_*, \mybm{w}) p(\mybm{w} | \mybm{X}, \mybm{r}) d\mybm{w} \\
& = \mathcal{N} ( \sigma_n^{-2} {\boldsymbol{\phi}_*}^{\top}\mybm{\tilde{K}}^{-1}\boldsymbol{\Phi}\mybm{r},  {\boldsymbol{\phi}_*}^{\top}\mybm{\tilde{K}}^{-1}\boldsymbol{\phi}_*) \\
& = \mathcal{N} ({\boldsymbol{\phi}_*}^{\top}\boldsymbol{\Sigma}_w\boldsymbol{\Phi}(\mybm{K} + \sigma_n^2\mybm{I}_M)^{-1}\mybm{r}, \\ 
& \qquad {\boldsymbol{\phi}_*}^{\top}\boldsymbol{\Sigma}_w\boldsymbol{\phi}_* - {\boldsymbol{\phi}_*}^{\top}\boldsymbol{\Sigma}_w\boldsymbol{\Phi}(\mybm{K} + \sigma_n^2\mybm{I}_M)^{-1}{\boldsymbol{\Phi}}^{\top}\boldsymbol{\Sigma}_w\boldsymbol{\phi}_*),
\end{split}
\end{equation}
where $\mybm{\tilde{K}} = \sigma_n^{-2}\boldsymbol{\Phi}{\boldsymbol{\Phi}}^{\top} + \boldsymbol{\Sigma}_w^{-1}$ and $\mybm{K} = {\boldsymbol{\Phi}}^{\top}\boldsymbol{\Sigma}_w\boldsymbol{\Phi}$. Here, note that it is more convenient to invert the $F \times F$ matrix $(\mybm{K} + \sigma_n^2\mybm{I}_F)$ than the $N \times N$ matrix $\mybm{\tilde{K}}$, whenever $F < N$. Furthermore, the feature space enters in the form ${\phi(\mybm{x})}^{\top}\boldsymbol{\Sigma}_w\phi(\myhat{\mybm{x}})$, where vectors $\mybm{x}$ and $\myhat{\mybm{x}}$ are either in the test or in the training dataset. At this point, let us define $k(\mybm{x},\myhat{\mybm{x}}) = {\phi(\mybm{x})}^{\top}\boldsymbol{\Sigma}_w\phi(\myhat{\mybm{x}})$ as the \emph{covariance function} or \emph{kernel}. Specifically, $k(\mybm{x},\myhat{\mybm{x}}) = {\phi(\mybm{x})}^{\top}\boldsymbol{\Sigma}_w\phi(\myhat{\mybm{x}})$ represents an inner product with respect to $\boldsymbol{\Sigma}_w$, equivalent to \mbox{$\boldsymbol{\psi}(\mybm{x}) \cdot \boldsymbol{\psi}(\myhat{\mybm{x}})$} when $\boldsymbol{\psi}(\mybm{x}) = \boldsymbol{\Sigma}_w^{1/2}\phi(\mybm{x})$ and $\boldsymbol{\Sigma}_w^{1/2}$ is such that $(\boldsymbol{\Sigma}_w^{1/2})^2 = \boldsymbol{\Sigma}_w$ (since $\boldsymbol{\Sigma}_w$ is positive definite). We can conclude that: if an algorithm is defined in terms of inner products in the input space, then it can be lifted into the feature space by replacing occurrences of inner products by the kernel, whenever it is more convenient to compute the kernel than the feature vectors themselves; this is also known as the \emph{kernel trick}~\cite{Rasmussen_2006}.\\

\noindent \textbf{2) The function-space view.} We can have exact correspondence with the weight-space view by using a GP modeling a distribution over functions. Formally: a GP is a collection of random variables, any finite number of which have a joint Gaussian distribution. Moreover, it is completely specified by the mean function and the \emph{covariance function} (or \emph{kernel}). We define the mean function and the \emph{covariance function} of process \mbox{$f(\cdot) \sim \mathcal{GP}(m(\mybm{x}), k(\mybm{x}, \myhat{\mybm{x}}))$} as 
\begin{equation}
\begin{split}
\label{eq:2.1}
m(\mybm{x}) &= \mathbb{E}[f(\mybm{x})] \\
k(\mybm{x}, \myhat{\mybm{x}}) &= \mathbb{E}[(f(\mybm{x}) - m(\mybm{x})){(f(\myhat{\mybm{x}}) - m(\myhat{\mybm{x}}))}^{\top}].
\end{split}
\end{equation}
Next, we consider the zero mean function, i.e., $m(\mybm{x}) = 0$, which is a very typical choice in the \ac{GP} literature~\cite{Rasmussen_2006}. In the Bayesian linear model of \eq{eq:1.1}, the \emph{prior} distribution is set to be Gaussian with zero mean and covariance matrix $\boldsymbol{\Sigma}_w$, i.e., \mbox{$\mybm{w} \sim \mathcal{N}(\mybm{0}, \boldsymbol{\Sigma}_w)$}. Thus, we can derive an example GP as:
\begin{equation}
\begin{split}
\label{eq:2.2}
m(\mybm{x}) &= {\phi(\mybm{x})}^{\top} \mathbb{E}[\mybm{w}] = \mybm{0} \\
k(\mybm{x}, \myhat{\mybm{x}}) &= {\phi(\mybm{x})}^{\top} \mathbb{E}[\mybm{w} \mybm{w}^{\top}] {\phi(\myhat{\mybm{x}})} = {\phi(\mybm{x})}^{\top} \boldsymbol{\Sigma}_w{\phi(\myhat{\mybm{x}})}.
\end{split}
\end{equation}
Assume the training dataset has $N$ observations, then vector \mbox{$\mybm{f} = [f(\mybm{x}_1), \dots, f(\mybm{x}_N)]^\top$} has a joint Gaussian distribution, i.e., \mbox{$\mybm{f}|\mybm{X} \sim \mathcal{N}(\mybm{0}, \mybm{K})$}, where the $N \times N$ covariance matrix $\mybm{K}$ can be computed evaluating the \emph{covariance function} or \emph{kernel} for the $N$ observations, i.e., $\mybm{K}_{ij} = {\phi(\mybm{x}_i)}^{\top} \boldsymbol{\Sigma}_w{\phi(\mybm{x}_j)}$ for $i,j = 1, \dots, N$. Given the noise \mbox{$\epsilon \sim \mathcal{N}(0, \sigma_n^2)$}, it follows from the i.i.d. assumption that a diagonal matrix $\sigma_n^2\mybm{I}_N$ must be added to $\mybm{K}$, as compared to the \mbox{noise-free} model in the GP literature~\cite{Rasmussen_2006}.
To make prediction for the test case $f(\mybm{x}_*) = {f}_*$ given $\phi(\mybm{x}_*) = \boldsymbol{\phi}_*$, we consider the joint Gaussian \emph{prior} distribution over functions
\begin{equation}
\label{eq:2.3}
    	 \begin{bmatrix}
           \mybm{r} \\
           {f}_* \\
         \end{bmatrix} = \mathcal{N} \bigg ( \mybm{0}, 
         \begin{bmatrix}
           & \mybm{K} + \sigma_n^2\mybm{I}_N & \mybm{k}_* \\
           & \mybm{k}_*^\top & k(\mybm{x}_*,\mybm{x}_*) \\
         \end{bmatrix} \bigg ),
\end{equation}
where we define the $N$-dimensional column vector $\mybm{k}_*$ such that the $i$-th element is equal to
${\phi(\mybm{x}_i)}^{\top} \boldsymbol{\Sigma}_w{\phi(\mybm{x}_*)}$. To derive the \emph{posterior} distribution over functions we need to condition the joint Gaussian \emph{prior} distribution over functions on the data, so that we get the key predictive equations of \acp{GP} for regression:
\begin{equation}
\begin{split}
\label{eq:2.4}
{f}_*&|\mybm{x}_*,\mybm{X},\mybm{r} \sim \mathcal{N}(\mu, \Sigma) \\
\mu & = \mybm{k}_*^\top[\mybm{K} + \sigma_n^2\mybm{I}_N]^{-1} \mybm{r} \\
\Sigma &= k(\mybm{x}_*, \mybm{x}_*) - \mybm{k}_*^\top[\mybm{K} + \sigma_n^2\mybm{I}_N]^{-1}\mybm{k}_*.
\end{split}
\end{equation}

\noindent In practice, the predictive mean $\mu$ is used as a point estimate for the function output, while the variance $\Sigma$ can be translated into uncertainty bounds (predictive error-bars) on this point estimate, thus making \acp{GP} for regression very appealing for \ac{MPC} applications (see~\cite{murray1999transient, leith2000nonlinear, solak2003derivative, kocijan2003predictive}).

For any set of basis functions in the feature space, we can compute the corresponding \emph{covariance function} or \emph{kernel}; conversely, for every (positive definite) \emph{covariance function} or \emph{kernel}, there exists a (possibly infinite) expansion in terms of basis functions in the feature space. As we show shortly, the choice of the kernel deeply affects the performance of a \ac{GP} for a given task, as much as the choice of the parameters (architecture, activation functions, learning rate, etc.) does for a neural network. Specifically, the \emph{hyperparameters} of the kernel must be set in order to optimize the \emph{marginal likelihood}, which is defined as follows:
\begin{equation}
\label{eq:2.5}
p(\mybm{r} | \mybm{X}) = \int p(\mybm{r} | \mybm{f},\mybm{X}) p(\mybm{f} | \mybm{X}) d\mybm{f}.
\end{equation}
Under the Gaussian assumption, the \emph{prior} distribution is Gaussian, \mbox{$\mybm{f} | \mybm{X} \sim \mathcal{N}(\mybm{0}, \mybm{K})$}, and the \emph{likelihood} is a factorized Gaussian, \mbox{$\mybm{r} | \mybm{f},\mybm{X} \sim \mathcal{N}(\mybm{f}, \sigma_n^2\mybm{I}_N)$}, thus \mbox{$\mybm{r} | \mybm{X} \sim \mathcal{N}(\mybm{0}, \mybm{K}+\sigma_n^2\mybm{I}_N)$}. Extensive derivation for the formulation of ${f}_*|\mybm{x}_*,\mybm{X},\mybm{r}$ and generalization to more that one test case can be found in~\cite{Rasmussen_2006}. 

\noindent Suppose we have $N_*$ observations in the test set, i.e., $(\mybm{X}_*,\mybm{r}_*)$, to make prediction for the test cases $f(\mybm{X}_*) = \mybm{f}_*$ given $\phi(\mybm{X}_*) = \boldsymbol{\Phi}_*$, we consider the joint Gaussian \emph{prior} distribution over functions
\begin{equation}
\label{eq:2.6}
    	 \begin{bmatrix}
           \mybm{r} \\
           \mybm{f}_* \\
         \end{bmatrix} = \mathcal{N} \bigg ( \mybm{0}, 
         \begin{bmatrix}
           & \mybm{K} + \sigma_n^2\mybm{I}_N & \mybm{K}_* \\
           & \mybm{K}_*^\top & \mybm{K}_{**} \\
         \end{bmatrix} \bigg ),
\end{equation}
where we define the $N \times N_*$ matrix $\mybm{K}_*$ similarly to $\mybm{k}_*$, such that 
$\mybm{K}_{*,ij} = {\phi(\mybm{x}_i)}^{\top} \boldsymbol{\Sigma}_w{\phi(\mybm{x}_{*,j})}$ for $i = 1, \dots, N$, $j = 1, \dots, N_*$, and $\mybm{x}_{*,j}$ is a column vector in $\mybm{X}_*$. 
Finally, we define the $N_* \times N_*$ matrix $\mybm{K}_{**}$ similarly to $k(\mybm{x}_*,\mybm{x}_*)$, such that 
$\mybm{K}_{**,ij} = {\phi(\mybm{x}_{*,i})}^{\top} \boldsymbol{\Sigma}_w{\phi(\mybm{x}_{*,j})}$ for $i,j = 1, \dots, N_*$, thus we get the key predictive equations of GPs for regression:
\begin{mdframed}
\begin{equation}
\begin{split}
\label{eq:2.7}
\mybm{f}_*&|\mybm{X}_*,\mybm{X},\mybm{r} \sim \mathcal{N}(\boldsymbol{\mu}, \boldsymbol{\Sigma}) \\
\boldsymbol{\mu} & = \mybm{K}_*^\top[\mybm{K} + \sigma_n^2\mybm{I}_I]^{-1} \mybm{r} \\
\boldsymbol{\Sigma} &= \mybm{K}_{**} - \mybm{K}_*^\top[\mybm{K} + \sigma_n^2\mybm{I}_I]^{-1}\mybm{K}_*.
\end{split}
\end{equation}
\end{mdframed}

\noindent \textbf{The choice of the kernel:} this choice deeply affects the performance of a GP for a given task, as it encodes the similarity between pairs of outputs in the function domain. There has been significant work on constructing base and composite kernels~\cite{duvenaud2013structure}. Common base kernels include the Squared Exponential (SE) kernel, the Rational Quadratic (RQ) kernel, and the Standard Periodic (SP) kernel, defined as:
\begin{equation}
\begin{split}
k_{\rm SE} (\mybm{x},\myhat{\mybm{x}}) & = \sigma_{\rm SE}^2 \exp(- ||\mybm{x} - \myhat{\mybm{x}} ||^2 / (2\ell_{\rm SE}^2)) \\
k_{\rm RQ} (\mybm{x},\myhat{\mybm{x}}) &= \sigma_{\rm RQ}^2 (1 + ||\mybm{x} - \myhat{\mybm{x}} ||^2 / (2\alpha_{\rm RQ}\ell_{\rm RQ}^2))^{-\alpha_{\rm RQ}} \\
k_{\rm SP} (\mybm{x},\myhat{\mybm{x}}) &= \sigma_{\rm SP}^2 \exp( - 2 \sin^2(\pi ||\mybm{x} - \myhat{\mybm{x}} || {\rm p}_{\rm SP}) / \ell_{\rm SP}^2).
\end{split}
\end{equation}

The properties of the functions under a GP with a SE kernel can display long range trends, where the length-scale $\ell_{\rm SE}$ determines how quickly a process varies with the inputs. The RQ kernel is derived as a scale mixture of SE kernels with different length-scales. The SP kernel is derived by mapping the two dimensional variable $(\cos(\mybm{x}); \sin(\mybm{x}))$ through the SE kernel. Derivations for the RQ and SP kernels are in~\cite{Rasmussen_2006}.

\begin{algorithm}[th]
	\caption{Pseudo-code for the basic routine
		\label{alg:alg 1}}
	\begin{algorithmic}[1] 
		\State Pre-training phase: find the optimal \emph{hyperparameters} $\boldsymbol{\theta}^{(0)}$ for the kernel $k(\cdot, \cdot)$, starting from $\boldsymbol{\theta}^{(s)}$ and minimizing the \emph{marginal likelihood} on the training dataset $\{(\mybm{x}_i , r_i )\}_{i=1}^{W}$
		\State Set $t = 1$
		\While{$t \le T-(N+N_*)$} 
		\State  Set $\mathcal{D}^{(t)} = (\mybm{X}^{(t)}, \mybm{r}^{(t)}) = \{(\mybm{x}_i , r_i )\}_{i=t-1+1}^{t-1+N}$
		\State  Set $\mathcal{D}^{(t)}_* = (\mybm{X}^{(t)}_*, \mybm{r}^{(t)}_*) = \{(\mybm{x}_i , r_i )\}_{i=t-1+N+1}^{t-1+N+N_*}$
		\State Training phase: find the optimal \emph{hyperparameters} $\boldsymbol{\theta}^{(t)}$ for the kernel $k(\cdot, \cdot)$, starting from $\boldsymbol{\theta}^{(0)}$ and minimizing the \emph{marginal likelihood} on the training dataset $(\mybm{X}^{(t)}, \mybm{r}^{(t)})$
		\State Get $(\boldsymbol{\mu}, \boldsymbol{\Sigma})$ via \eq{eq:2.7} given the test set $(\mybm{X}^{(t)}_*, \mybm{r}^{(t)}_*)$
		\State Compute ${\rm RMSE}^{(t)}_* = \sqrt{(\sum_{i = 1}^{N_*} e_i^2)/ N_*}, \ \mybm{e} = \mybm{r}^{(t)}_* - \boldsymbol{\mu}$
		\State Set $t = t+1$
		\EndWhile
	\end{algorithmic}
\end{algorithm}

Note that valid kernels (i.e., those having a \mbox{positive-definite} covariance function) are closed under the operators $+$ and $\times$. This allows one to create more representative (and composite) kernels from \mbox{well-understood} basic components, according to the following key rules~\cite{duvenaud2013structure}:
\begin{itemize}
\item Any subexpression\footnote{Subexpression refers to any valid kernel family, either basic or composite.} $\mathcal{P}$ can be replaced with $\mathcal{P}+\mathcal{B}$, where $\mathcal{B}$ is any base kernel family. 
\item Any subexpression $\mathcal{P}$ can be replaced with $\mathcal{P} \times \mathcal{B}$, where $\mathcal{B}$ is any base kernel family. 
\item Any base kernel $\mathcal{B}$ can be replaced with any other base kernel family  $\mathcal{B}'$. 
\end{itemize}

In time series, summing kernels can express superpositions of different processes, operating at different scales, whereas multiplying kernels may be a way of converting global data properties onto local data properties. From here on, we will use \mbox{one-dimensional} kernels in the form ${\rm RQ}\times {\rm SP}$ with period ${\rm p}_{\rm SP}$, which correspond to a local \mbox{quasi-periodic} structure in the data, with noise operating at different scales. Note that kernels over multidimensional inputs can be constructed via the operators $+$ and $\times$ over individual dimensions. Next, we consider models based on zero-mean \acp{GP} for the runtime \mbox{multi-step} ahead forecasting of time series, with application to a) Harvested Energy Profile $H(t)$~(defined in \secref{sub:harvested_energy}) and b) Traffic Load $L(t)$~(\secref{sub:traffic_load}). \\ 

\noindent \textbf{The basic routine for prediction:} we use models based on \mbox{zero-mean} \acp{GP} for the runtime forecasting of time series, with application to $H(t)$ and $L(t)$, $t = 1, \dots, T$. The strong daily seasonality of the data is evident for both time series, as well as the presence of noise at different scales. Therefore, we define composite kernels for $H(t)$ and $L(t)$ in the form ${\rm RQ} \times {\rm SP}$ with period ${\rm p}_{\rm SP}$, i.e.,
\begin{mdframed}
\begin{equation}
\begin{split}
\label{eq:kernel}
k(x, \myhat{x}) = & \sigma^2\exp( - 2 \sin^2(\pi d {\rm p}_{\rm SP}) / \ell_{\rm SP}^2) \\
& \times (1 + d^2 / (2\alpha_{\rm RQ}\ell_{\rm RQ}^2))^{-\alpha_{\rm RQ}}
\end{split}
\end{equation}
\end{mdframed}
where $\sigma = \sigma_{\rm RQ} \sigma_{\rm SP}$ and $d = |x -\myhat{x}|$ is the Euclidean distance between inputs. At this point, the \emph{hyperparameters} of the kernel must be set in order to optimize the \emph{marginal likelihood}, which is defined in \eq{eq:2.5}, and here implemented using the toolbox of~\cite{toolbox}. 
For compactness, we aggregate the \emph{hyperparameters} of the kernel in the initialization set $\boldsymbol{\theta}^{(s)} = \{ \sigma, {\rm p}_{\rm SP}, \ell_{\rm SP}, \alpha_{\rm RQ}, \ell_{\rm RQ} \}$. Here, we opt for $\sigma = 1$, ${\rm p}_{\rm SP} = 24$, and select the free parameters ($\ell_{\rm SP}$, $\alpha_{\rm RQ}$, $\ell_{\rm RQ}$) via a grid search, scanning combinations in the range $[10^{-2},10^2]$. 
To model the strong daily seasonality in the data, we also opt for a prior distribution on the period ${\rm p}_{\rm SP}$, which is a delta function, i.e., $\delta({\rm p}_{\rm SP} - 24) = 1$ if and only if ${\rm p}_{\rm SP} = 24$, so that we treat the period ${\rm p}_{\rm SP}$ as a constant, excluding it from the optimization (see~\cite{toolbox}).

\begin{table}[t]
	\centering
	\begin{subtable}{.5\textwidth}
		\centering
		\begin{tabular}{|l|l|l|l|l|}
			\hline
			& $N_*=1$  & $N_*=2$ & $N_*=12 $ & $N_*=24$ \\ \hline
			$S=1$ & 0.0119 & 0.0170 & 0.0385 & 0.0512 \\ \hline
			$S=T$ & 0.0116 & 0.0166 & 0.0383 & 0.0511 \\ \hline
		\end{tabular}
		\caption{Average ${\rm RMSE}^{(t)}_*$ for $H(t)$.} \label{table:A}
	\end{subtable}
	\quad
	\begin{subtable}{.5\textwidth}
		\centering
		\begin{tabular}{|l|l|l|l|l|}
			\hline
			& $N_*=1$  & $N_*=2$ & $N_*=12$ & $N_*=24$ \\ \hline
			$S=1$ & 0.0389 & 0.0464 & 0.0670 & 0.0740 \\ \hline
			$S=T$ & 0.0415 & 0.0483 & 0.0671 & 0.0743 \\ \hline
		\end{tabular}
		\caption{Average ${\rm RMSE}^{(t)}_*$ for $L(t)$.}
		
		\label{table:B}
	\end{subtable}
	\caption{Average ${\rm RMSE}^{(t)}_*$.}
\end{table}

Algorithm~\ref{alg:alg 1} describes the basic routine for the \mbox{pre-training}, training and forecasting phases for both zero-mean GPs, i.e., the same basic reasoning holds for $H(t)$ and $L(t)$, where $\mybm{x}_t$ refers to time $t$ and $r_t$ refers to either $H(t)$ or $L(t)$, at time $t$. Also, we assume that we can access the $N$ values in the training dataset, and we wish to predict the $N_*$ values in the test set, where $\mathcal{D}^{(t)} = (\mybm{X}^{(t)}, \mybm{r}^{(t)})$ refers to the training dataset and $\mathcal{D}^{(t)}_* = (\mybm{X}^{(t)}_*, \mybm{r}^{(t)}_*)$ refers to the test set, at time $t$, respectively.
According to the pre-training phase, we first have to find the optimal \emph{hyperparameters} $\boldsymbol{\theta}^{(0)}$ for the kernel $k(\cdot, \cdot)$, starting from $\boldsymbol{\theta}^{(s)}$ and minimizing the \emph{marginal likelihood} on the training dataset $\{(\mybm{x}_i , r_i )\}_{i=1}^{W}$, where we set $W = N$. 
Note that $\boldsymbol{\theta}^{(0)}$ will serve as initialization for the optimal \emph{hyperparameters} $\boldsymbol{\theta}^{(t)}$ at each step of the online forecasting routine, as the optimal \emph{hyperparameters} $\boldsymbol{\theta}^{(t)}$ are found over the training dataset $(\mybm{X}^{(t)}, \mybm{r}^{(t)})$, which changes at each step of the online forecasting routine. Assuming Gaussian noise with variance $\sigma_n^2$, thus Gaussian likelihood, it follows that we can perform exact inference. To do it, we use the Conjugate Gradients (CG) optimization tool implemented in toolbox~\cite{toolbox}. 
We get $(\boldsymbol{\mu}, \boldsymbol{\Sigma})$ via \eq{eq:2.7} given the test set $(\mybm{X}^{(t)}_*, \mybm{r}^{(t)}_*)$ with $N_*$ test cases, at time $t$. Finally, we derive the Root Mean Square Error (RMSE) ${\rm RMSE}^{(t)}_*$ over the $N_*$ test cases, starting from residuals $\mybm{e}$, at time $t$, and iterating the procedure (except for the pre-training phase) up to time $T-(N+N_*)$. For the numerical results, the training phase is performed once every $S$ steps.\\

\noindent \textbf{Numerical results:} now, we assess the proposed scheme for the runtime \mbox{multi-step} ahead forecasting of time series $H(t)$ and $L(t)$, where we track ${\rm RMSE}^{(t)}_*$ over the $N_*$ test cases, given $N_* = 1, 2, 12, 24$. Here, we set the time step to one hour, $N = 24 \times 14 = 336$~hours (i.e., two weeks of data), $T = 24 \times 60 = 1440$~hours (i.e., two months of data), $\sigma_n = 10^{-5}$, and we recall that $W = N$. This choice of parameters is valid for both time series, as well as the use of the kernel $k(\cdot,\cdot)$ in \eq{eq:kernel}, whereas the \emph{hyperparameters} differ, depending on the nature of data.

\begin{figure}[t]
	\centering
	\begin{subfigure}[t]{\columnwidth}
		\centering
		\includegraphics[width=\columnwidth]{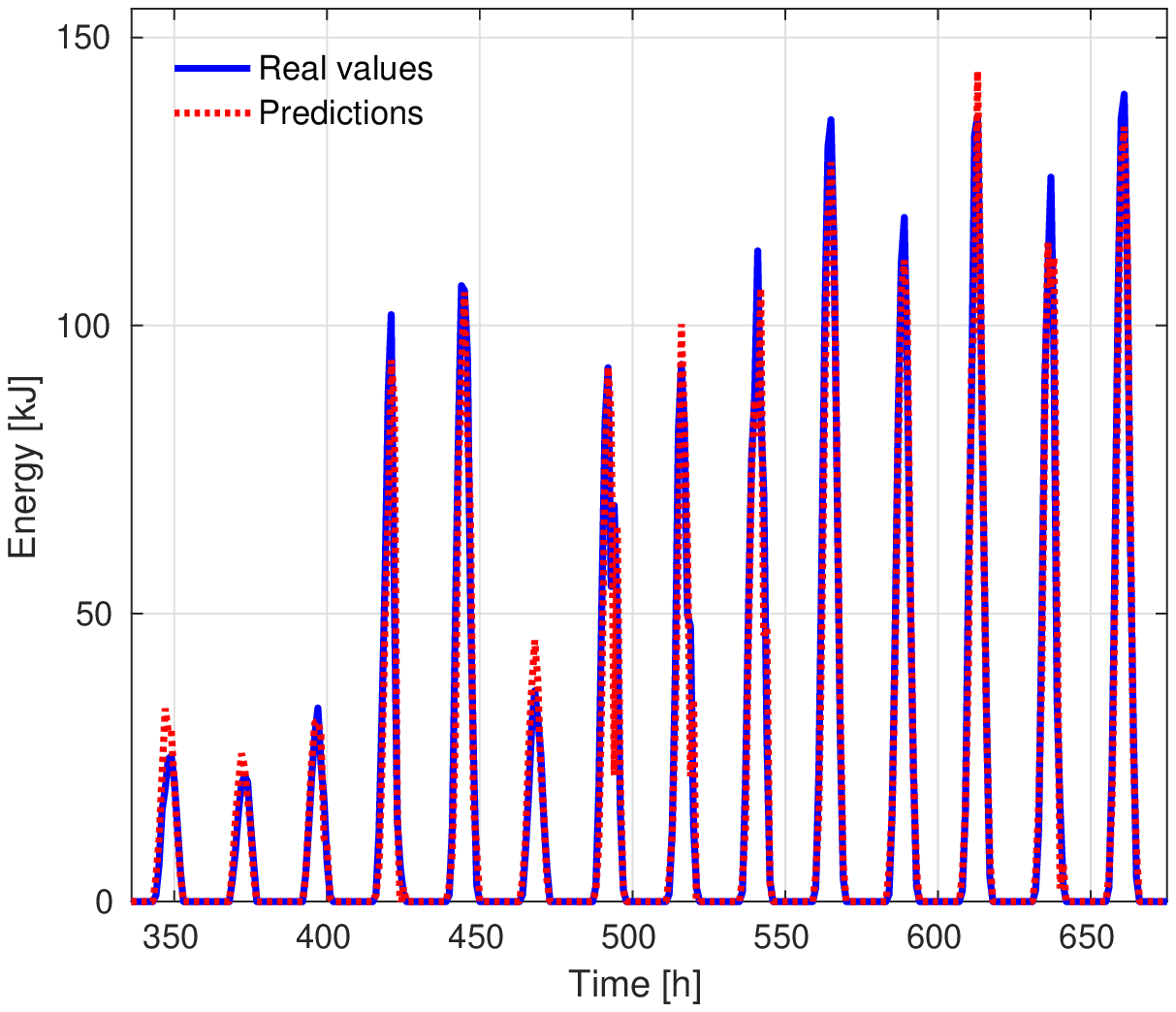}
		\caption{One-step predictive mean value for $H(t)$.}\label{figure:Fig_A_all}	
	\end{subfigure}
	\quad
	\begin{subfigure}[t]{\columnwidth}
		\centering
		\includegraphics[width=\columnwidth]{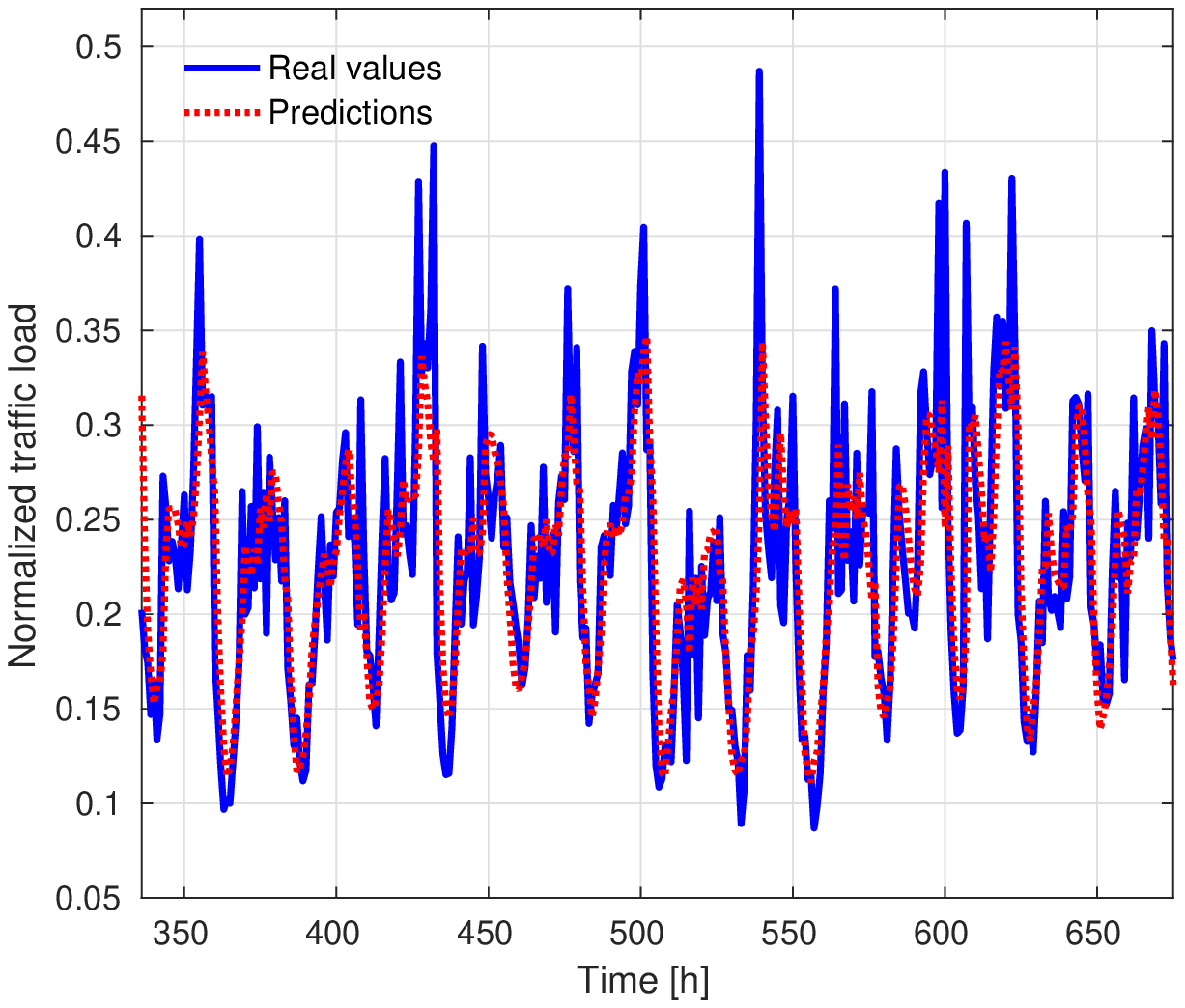}
		\caption{One-step predictive mean value for $L(t)$.}\label{figure:Fig_B_all}	
	\end{subfigure}
	\centering
	\caption{One-step online forecasting for one-month of data.}
	\label{figure:Fig_AB_all}
\end{figure}

\begin{figure}[t]	
	\centering
	\begin{subfigure}[t]{\columnwidth}
		\centering
		\includegraphics[width=\columnwidth]{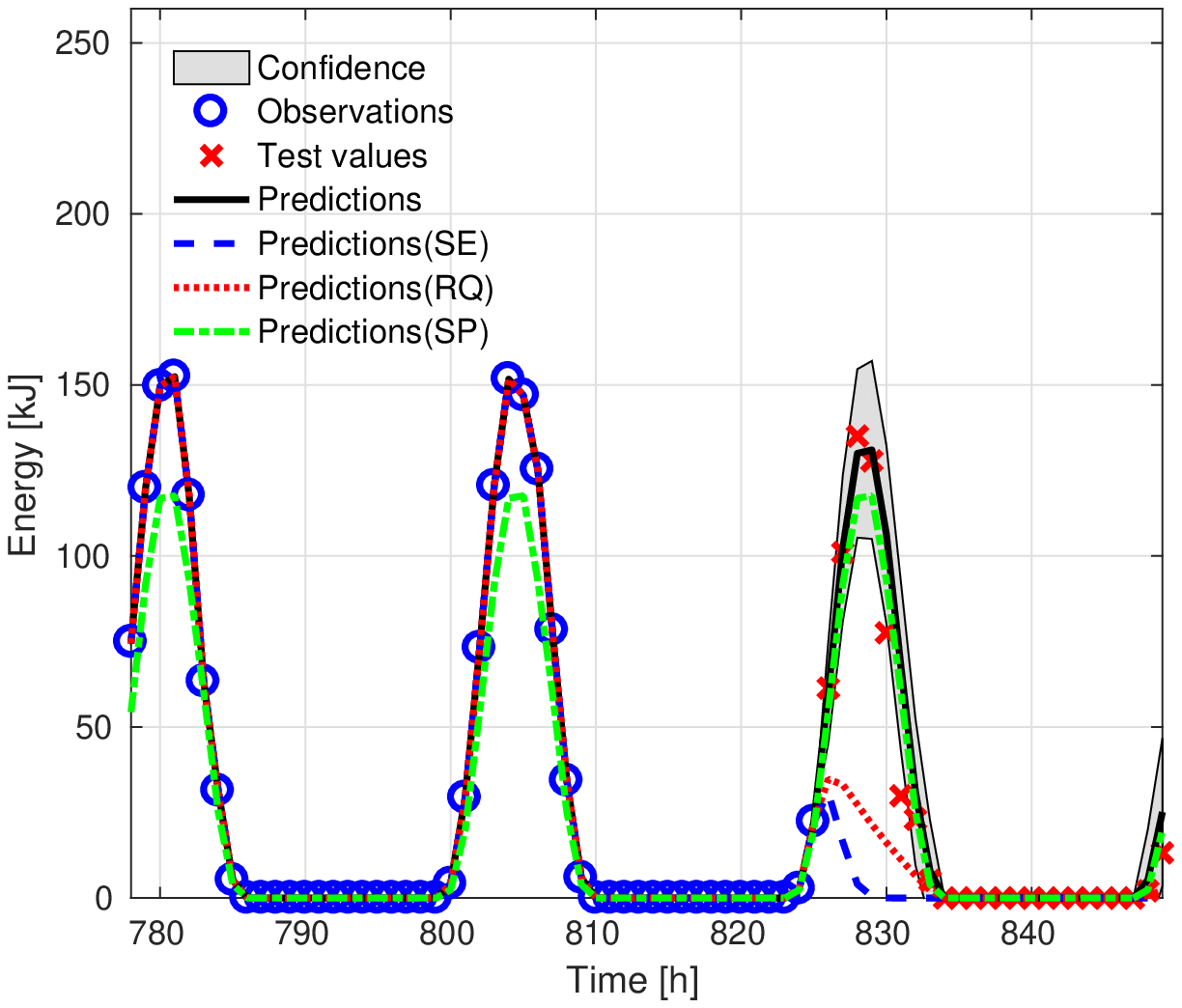}
		\caption{Multi-step predictive mean value for $H(t)$.}
		\label{fig:multi_h}	
	\end{subfigure}
	\quad
	\begin{subfigure}[t]{\columnwidth}
		\centering
		\includegraphics[width=\columnwidth]{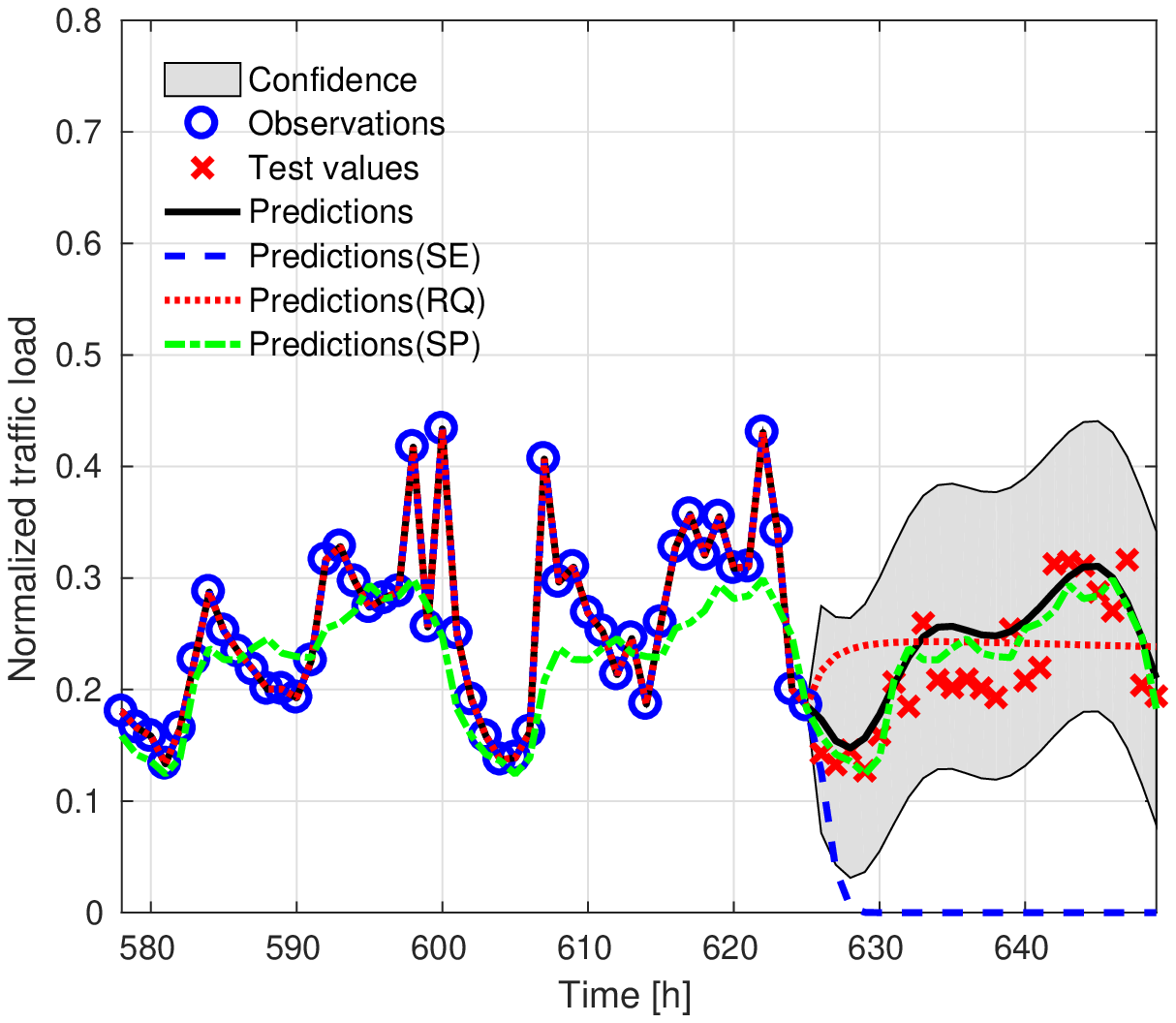}
		\caption{Multi-step predictive mean value for $L(t)$.}
		\label{fig:multi_l}	
	\end{subfigure}
	\caption{Multi-step prediction with different kernels.}
	\label{fig:multi}
\end{figure}

In Table~\ref{table:A} and Table~\ref{table:B} we show the average RMSE for $H(t)$ and $L(t)$, computed evaluating the mean of the RMSE measures up to time $T-(N+N_*)$, where we track ${\rm RMSE}^{(t)}_*$ over the $N_*$ test cases, given $N_* = 1, 2, 12, 24$. Also, as we perform the training phase once every $S$ steps, we compare the numerical results when $S = 1$ and $S = T$, i.e., when we \mbox{re-optimize} the \mbox{free-parameters} at each step of the online forecasting routine, or just once every $T$ steps, at time $t=1$. In general, the average ${\rm RMSE}^{(t)}_*$ decreases as we increase the $N_*$ test cases up to $24$, which corresponds to one day into the future. However, the worst performance is $0.0743$, which is still rather small if we consider that both time series are normalized in $[0,1]$ prior to processing. Also, predictions for $H(t)$ (Fig.~\ref{figure:Fig_A_all}) are more precise than predictions for $L(t)$ (Fig.~\ref{figure:Fig_B_all}), and this is due to the nature of the data, given that we use the same kernel for both time series. In fact, values in $H(t)$ (Fig.~\ref{figure:Fig_A_all}) follow a more regular behavior than those in $L(t)$ (Fig.~\ref{figure:Fig_B_all}), with \mbox{quasi-periodic} streams of zero values corresponding to zero solar energy income during the night. These \mbox{quasi-periodic} streams of zero values help reinforcing prediction, while allowing for a higher confidence at nighttime (see Fig.~\ref{fig:multi_h}). 
Finally, tuning parameter $S$ explains the impact of \mbox{re-optimizing} the \emph{hyperparameters} according to the most recent history (i.e., two weeks of data), but with a longer execution time. Numerical results suggest that tuning parameter $S$ could be reasonable when data exhibit multiple strong local behaviors rather than just a strong daily seasonality, and the kernel has to adapt to these. However, $S=1$ could not be the obvious, optimal choice (see Table~\ref{table:A}).

In Fig.~\ref{figure:Fig_A_all} and Fig.~\ref{figure:Fig_B_all} we show real values and predictions for two weeks of data, where we track the \mbox{one-step} predictive mean value at each step of the online forecasting routine. 
The strong daily seasonality is evident, as well as the \mbox{quasi-periodic} structure in data with noise operating at different scales. Note that predictions for $H(t)$ (Fig.~\ref{figure:Fig_A_all}) are more accurate than those for $L(t)$ (Fig.~\ref{figure:Fig_B_all}), and this result can be confirmed by comparing the average ${\rm RMSE}^{(t)}_*$ in Tables~\ref{table:A} and~\ref{table:B} for $N_* = 1$. However, predictions are still quite far from real values when some unusual events occur, see, for example, the low solar energy income within hours $456$ and $480$ (sixth peak from the left), in Fig.~\ref{figure:Fig_A_all}, or the sudden peaks in the traffic load profile of Fig.~\ref{figure:Fig_B_all}, which are very \mbox{day-specific}.

In Figs.~\ref{fig:multi_h} and~\ref{fig:multi_l} we show real and predicted values for three days of data, i.e., the last two days of the training dataset, and $24$ hours for the test set, plotting the \mbox{multi-step} predictive mean value with $N_*=24$. Here, we compare the use of the kernel $k(\cdot,\cdot)$ in \eq{eq:kernel} with common base kernels from the literature, such as the popular Squared Exponential (SE) kernel, the Rational Quadratic (RQ) kernel, and the Standard Periodic (SP) kernel, see~\eq{eq:kernel}. Also, we compare the use of the kernel $k(\cdot,\cdot)$ in \eq{eq:kernel} in terms of generalization capabilities over the training dataset and the test set, i.e., we perform forecasting over the training dataset and the test set, after the optimization of the \emph{hyperparameters} given the observations. Note that the proposed kernel (solid line) shows the best performance in terms of forecasting, since composite kernels are more representative than base ones. Specifically, the RMSE is close to zero over the training dataset (due to the fact that we set $\sigma_n = 10^{-5}$, i.e., $\sigma_n \neq 0$), and this result also holds for both the SE and RQ cases. However, the generalization capabilities over the test set are quite limited for SE and RQ. In fact, these base kernels have limited expressive power, and simply act like smoothers. Finally, the SP kernel succeeds in recovering the strong daily seasonality in the data, but it fails to model noise at different scales. Again, its expressive power is quite limited, with respect to our proposed kernel in \eq{eq:kernel}.


\subsection{Predictive and time-adaptive energy allocation}
\label{sub:model_predictive_control}



In general, an MPC framework can be divided into three blocks: i) inputs, ii) MPC controller and iii) real system~\cite{maciejowski2013fault}. The first block is the prediction model (see Section~\ref{sub:prediction}). The MPC solves a control problem  at runtime (see below). Finally, the real system block receives the optimal actions from the MPC controller and behaves accordingly. \\

\noindent \textbf{Notation:} using a standard notation, the system to be controlled is described by means of a \mbox{discrete-time} model:
\begin{equation}
\label{eq:mpc1}
\bm{Z}_{t+1} = \bm{Z}_{t} + \bm{U}_{t} + \bm{W}_{t},
\end{equation}
where $t$ is the current time slot. The $M \times n_s$ matrix $\bm{Z}_t$ with elements $z_{kn}$ denotes the {\it system state}, representing for each BS \mbox{$n \in \mathcal{S}$} the energy buffer level for time slots \mbox{$k = t, t+1, \dots,  t+M-1$}, were $M$ is the optimization horizon (we use $M=24$~hours), which corresponds to $N_{*}$ in the previous subsection. The $M \times n_s$ matrix $\bm{U}_t$ with elements $u_{kn}$ denotes the {\it control} matrix, representing the amount of energy that each BS $n$ shall either transfer or receive (depending on the sign of $u_{kn}$) in time slot $k = t, \dots, t+M-1$. The $M \times n_s$ matrix $\bm{W}_t$ models the system disturbances, i.e., the stochastic behavior of the forecast profiles (harvested and consumed energy), with: 
\begin{equation}
\label{eq:mpc2}
\bm{W}_{t} \sim \mathcal{N}(\bm{\widebar{W}}_{t}, \Sigma_{\bm{W}_{t}}),
\end{equation}
where $\bm{\widebar{W}}_{t}$ and $\Sigma_{\bm{W}_{t}}$ contain the mean and variance of the forecast estimates, respectively. Eqs.~\eqsimple{eq:mpc1} and \eqsimple{eq:mpc2} relate to the problem setup of Section~\ref{sub:storage_units} as follows: symbol $\bm{Z}_t$ contains the buffer state for all \acp{BS}, i.e., $z_{kn} = B_n(k)$, $\bm{U}_t$ is the control, which corresponds to the amount of energy to transfer, i.e., $u_{kn} = T_n(k)$, and $\bm{W}_t$ contains the exogenous processes, i.e., \mbox{$w_{kn} = H_n(k)-O_n(k)$}. Note that processes $H_n(k)$ and $O_n(k)$ are statistically characterized through the prediction framework of Section~\ref{sub:GP}, and their difference is still a Gaussian r.v. (in fact, $O_n(k)$ is derived from $L_n(k)$ through a linear model, and as such is still Gaussian distributed). 
Following~\cite{wang2016stochastic}, due to the stochastic nature of \eq{eq:mpc2}, the system state $\bm{Z}_t$ should also be written in a probabilistic way:
\begin{equation}
\label{eq:mpc3}
\bm{Z}_t \sim \mathcal{N}(\bm{\widebar{Z}}_t, \Sigma_{\bm{Z}_t}),
\end{equation}
where $\bm{\widebar{Z}}_t$ and $\Sigma_{\bm{Z}_t}$ are the mean and the variance of $\bm{Z}_t$, respectively.\\

\noindent \textbf{Objective functions:} the goal of the MPC controller is to determine the amount $u_{kn}$ that each \ac{BS} $n$ should either transfer or receive in time slots $k=t,\dots,t+M-1$. If $u_{kn} >0$, \ac{BS} $n$ acts as a source in slot $k$, whereas if $u_{kn}<0$ it acts as an energy consumer. A first cost function tracks the total amount of energy that is exchanged among \acp{BS}:
\begin{equation}
\label{eq:mpc_obj1}
f_{1}^{MPC}(\bm{U}_t) = \sum_{k=t}^{t+M-1} \sum_{n=1}^{n_s} \left (u_{kn}\right)^2.
\end{equation}
For the second objective, we use the reference threshold $\bref$, see Section~\ref{sub:storage_units}. The MPC controller tries to increase the \ac{BS} energy buffer levels, while avoiding that only a few of the consumers receive energy. To achieve this, a second cost function weighs how close to $\bref$ the buffers get:
\begin{equation}
\label{eq:mpc_obj2}
f_{2}^{MPC}(\bm{Z}_t, \bref) = \sum_{k=t}^{t+M-1} \sum_{n=1}^{n_s} (z_{kn}-\bref)^2,
\end{equation}
where $z_{kn}$ is the energy buffer level of \ac{BS} $n$ in time slot $k$. \\

\noindent \textbf{Control problem:} once the inputs are stated, a \mbox{finite-horizon} \mbox{multi-objective} optimization problem is formulated:

\begin{mdframed}[innertopmargin=0pt]
\begin{subequations}
	\label{eq:mpc_formulation}
	\begin{align}
	& \underset{\bm{U}_t}{\text{min}}
	& & \mathrm{} \mathbb{E} \left[ \alpha f_{1}^{MPC}(\bm{U}_t) + (1 - \alpha) f_{2}^{MPC}(\bm{Z}_t, \bref) \right] \\
	& \text{subject to:}
	& &  \bm{Z}_{t} \sim \mathcal{N}(\bm{\widebar{Z}}_t, \Sigma_{\bm{Z}_t}), \\
	& \text{}
	& &  \bm{W}_t \sim \mathcal{N}(\bm{\widebar{W}}_t, \Sigma_{\bm{W}_t}), \\
	& \text{}
	& & \blow \leq z_{kn} \leq B_{max},	\\
	& \text{}
	& & u_{kn}^\min \leq u_{kn} \leq u_{kn}^\max, \\
	& \text{}
	& & \textrm{with: } k=t,t+1,\dots,t+M-1 \nonumber
	\end{align}
\end{subequations}
\end{mdframed}

\noindent where $\alpha \in [0,1]$ is a weight to balance the relative importance of the two cost functions. $\blow$ and $\bmax$ are the energy buffer limitations defined in Section~\ref{sub:storage_units}. Finally, the fourth constraint defines the amount of energy that each BS $n \in n_s$ can exchange and depends on the system state, i.e., the energy buffer level, expected harvested energy and expected traffic load. The system state dictates the limits of the control action.

For any fixed value of $\alpha$, and since the optimization problem must be solved at runtime, it is strongly preferable to choose a convex optimization formulation such as \eq{eq:mpc_formulation}, which can be solved through standard techniques. Here, we have used the \texttt{CVX} tool~\cite{grant2008cvx} to obtain the optimal solution $\bm{U}_t^*=[u_{kn}^*]$, which dictates the optimal amount of energy that each BS $n \in n_s$ shall either provide or receive in time slot $k = t, \dots, t+M-1$.\\

\noindent \textbf{Optimization algorithm:} the MPC controller performs as follows~\cite{zhang2017optimal}:
\begin{enumerate}
\item \textbf{Step 1:} at the beginning of time slot $k$, the system state is obtained, that is energy buffer levels for all \acp{BS}, the  harvested energy and traffic load forecasts for the next $M$ hours (the optimization horizon).
\item \textbf{Step 2:} the control problem in \eq{eq:mpc_formulation} is solved yielding a sequence of control actions over the horizon $M$.
\item \textbf{Step 3:} only the first control action is performed and the system state is updated upon implementing the required energy transfers.
\item \textbf{Step 4:} Forecasts are updated and the optimization cycle is repeated from \textbf{Step 1}.
\end{enumerate}

\subsection{Energy scheduling for the current time slot}
\label{sub:convex_optimization}


With the \ac{MPC} of the previous subsection, we compute the amount of energy that each \ac{BS} should either provide (in case the \ac{BS} acts as a source) or receive (if the \ac{BS} is a consumer). Note that the \ac{BS} role (source or consumer) is also a result of the optimization process. In this section, we solve the energy allocation problem between energy sources and energy consumers, i.e., which source will transfer energy to which consumer and in which amount. Note that this also depends on the distribution losses between them and, in turn, on the electrical \ac{PPG} topology.\\

\noindent \textbf{Notation:} we use indices $i$ and $j$ to respectively denote an arbitrary \ac{BS} source and an arbitrary \ac{BS} consumer. \mbox{$\mathcal{Y}_s = \{1, \dots i, \dots, I\}$} and \mbox{$\mathcal{Y}_c = \{1, \dots j, \dots, J\}$} are the set of sources and consumers, respectively. With $e_{ij}$ we mean the energy available from the energy source $i \in \mathcal{Y}_s$ to the energy consumer $j \in \mathcal{Y}_c$, in matrix notation we have $\bm{E}= [e_{ij}]$. Note that $e_{ij}$ is the energy that would be available at the consumer BS $j$ and, in turn, it depends on $i$, $j$ and on the distribution losses between them, i.e., on the total distance that the energy has to travel, see Section~\ref{electrical_grid}. Vector $\bm{d}$, with elements $d_j$, represents the energy demand from each consumer $j \in \mathcal{Y}_c$. $g_{ij}$ represents the number of hops in the energy routing topology between source $i \in \mathcal{Y}_s$ and consumer $j \in \mathcal{Y}_c$, in matrix notation we have $\bm{G} = [g_{ij}]$. Also, we assume that all hops have the same physical length. Finally, with $y_{ij} \in [0,1]$ we mean the fraction of $e_{ij}$ that is allocated from source $i \in \mathcal{Y}_s$ to  consumer $j \in \mathcal{Y}_c$, in matrix notation $\bm{Y} = [y_{ij}]$.	\\

\noindent \textbf{Objective functions:} as a first objective, we seek to minimize the difference between the amount of energy offered by the BS sources $i \in \mathcal{Y}_s$ and that transferred to the BS consumers $j \in \mathcal{Y}_c$. This amounts to fulfill, as much as possible, the consumers' energy demand. At time $t$, the energy that can be drained from a source $i$ is $u_{ti}$. Now, if we consider the generic consumer $j$, the maximum amount of energy that $i$ can provide to $j$ is \mbox{$e_{ij}= u_{ti} a(g_{ij})$}, where $a(g_{ij}) \in [0,1]$ is the attenuation coefficient between $i$ and $j$, due to the power loss. We thus write a first cost function as: 
\begin{equation}
\label{eq:obj1}
f_{1}(\bm{Y}, \bm{E}, \bm{d}) = \sum_{j=1}^{J} \left (\sum_{i=1}^{I} y_{ij}e_{ij} - d_j \right )^2 \, ,
\end{equation}
where $i \in \mathcal{Y}_s$ and $j \in \mathcal{Y}_c$. Due to the existence of a single path between any source and consumer pair and due to the fact that each power link can only be used for a single transfer operation at a time, a desirable solution shall: i) pick source and consumer pairs $(i,j)$ in such a way that the physical distance ($g_{ij}$) between them is minimized and ii) achieve the best possible match between sources and consumers, i.e., use source $i$, whose available energy is the closest to that required by consumer $j$. In other words, we would like $y_{ij}$ to be as close as possible to $1$. If this is infeasible, multiple sources will supply the consumer. Minimizing the following cost function, the number of hops $g_{ij}$ between sources and consumers is kept small and we favor solutions with $y_{ij} \to 1$:
\begin{equation}
\label{eq:obj2}
f_{2}(\bm{Y}, \bm{G}) = \sum_{i=1}^{I} \left (\sum_{j=1}^{J} -\exp \left (\frac{y_{ij}}{g_{ij}} \right ) \right ) \, ,
\end{equation}
that is, with this cost function we are looking for a sparse solution (i.e., a small number of sources with $y_{ij}$ close to $1$). Note that when $y_{ij} \to 1$ and $g_{ij}$ is minimized, the argument $y_{ij} / g_{ij}$ is maximized and the negative exponential is minimized. Also, the exponential function was picked as it is convex, but any increasing and convex function would do.\\

\noindent \textbf{Solution through convex optimization:} at each time slot $t$, each \ac{BS} $n$ updates its buffer level $B_n(t)$, using either \eq{eq:buffer_1} or \eq{eq:buffer_2} (note that $B_n(t-1)$, $H_n(t-1)$, $O_n(t-1)$, $T_n(t-1)$ and $\theta_n(t-1)$ are all known in slot $t$, see Section~\ref{sec:systemModel}). The \ac{MPC} problem of Section~\ref{sub:model_predictive_control} is solved, and in the current time slot $t$, \ac{BS} $n$ acts as a source if $u_{tn}>0$ and as a consumer if $u_{tn}<0$. Each source $i$ evaluates $e_{ij}$ for all $j \in \mathcal{Y}_c$ through $e_{ij}= u_{ti} a(g_{ij})$ and each consumer $j$ sets its energy demand as \mbox{$d_j = u_{tj}$}. At time $t$, using \eq{eq:obj1} and \eq{eq:obj2}, the following optimization problem is formulated:
\begin{mdframed}
\begin{subequations}
	\label{eq:opt_prob}
	\begin{align}
	& \underset{\bm{Y}}{\text{min}}
	& & \mathrm{} \beta f_{1}(\bm{Y}, \bm{E}, \bm{d}) + (1 - \beta) f_{2}(\bm{Y}, \bm{G})\\
	& \text{subject to:}
	& &  0 \leq y_{ij} \leq 1 , \qquad \forall i \in \mathcal{Y}_s, \forall j \in \mathcal{Y}_c, \\
	& \text{}
	& &  \sum_{j=1}^{J} y_{ij} \leq 1 , \qquad \, \forall i \in \mathcal{X}_s,
	\end{align}
\end{subequations}
\end{mdframed}
where $\beta \in [0,1]$ is a weight used to balance the relative importance of the two cost functions. The first constraint represents the fact that $y_{ij}$ is a fraction of the available energy $e_{ij}$ from source $i$, and the second constraint means that the total amount of energy that a certain source $i$ transfers to consumers $j = 1, \dots, J$ cannot exceed the total amount of available energy at this source.

For any fixed value of $\alpha$, \eq{eq:opt_prob} is a convex minimization problem which can be solved through standard techniques. In this paper, we have used the \texttt{CVX} tool~\cite{grant2008cvx} to obtain the optimal solution $\bm{Y}^*=[y_{ij}^*]$, which dictates the optimal energy fraction to be allocated from any source $i$ to any consumer $j$.\\

\noindent \textbf{Solution through the Hungarian method:} the energy distribution problem from sources to consumers can also be modeled as an assignment problem, where each source $i \in \mathcal{Y}_s$ has to be {\it matched} with a consumer $j \in \mathcal{Y}_c$. This approach can be solved through the {\it Hungarian method}~\cite{kuhn1955hungarian}, an algorithm capable of finding an optimal assignment for a given square $A\times A$ cost matrix, where $A = \max(I,J)$. An assignment is a set of $A$ entry positions in the cost matrix, no two of which lie in the same row or column. The sum of the $A$ entries of an assignment is its cost. An assignment with the smallest possible cost is referred to as {\it optimal}. 
Let $\bm{C}=[c_{ij}]$ be the cost matrix, where rows and columns respectively correspond to sources $i$ and consumers $j$. Hence, $c_{ij}$ is the cost of assigning the \mbox{$i$-th} source to the \mbox{$j$-th} consumer and is obtained as follows:
\begin{equation}
\label{eq:cost_hungarian}
c_{ij} = \beta (e_{ij} - d_j)^2 + (1 - \beta) \left ( -\exp\left(\frac{1}{g_{ij}}\right)  \right ) \, ,
\end{equation}
where $\beta \in [0,1]$, the first term weighs the quality of the match ($d_j$ should be as close as possible to $e_{ij}$) and the second the quality of the route. To ensure the cost matrix is square, additional rows or columns are to be added when the number of sources and consumers differs. As typically assumed, each element in the added row or column is set equal to the largest number in the matrix.

The main difference between the optimal solution found by solving the convex optimization problem (\eq{eq:opt_prob}) and that found by the Hungarian method is that the latter returns a \mbox{one-to-one} match between sources and consumers, i.e., each consumer can only be served by a single source. On the other hand, for any given consumer the convex solution {\it also allows the energy transfer from multiple sources}.

\subsection{Energy Routing Strategy}
\label{sec:scheduling}

Now, we describe how the energy allocation $y_{ij}$ is implemented over time. The algorithm that follows is executed at the beginning of each time slot, when a new allocation matrix $\bm{Y}^*$ is returned by the solver of Section~\ref{sub:convex_optimization}. Each hour is further split into a number of mini slots. Given a certain maximum transmission energy capacity $e_{\max}$ for a power link in a mini slot, the required number of mini slots to deliver a certain amount of power $y_{ij}e_{ij}$ between source $i$ and consumer $j$ is obtained as $n_{ij} = \lceil y_{ij}e_{ij}/e_{\max} \rceil$.

Since each power link can only be used for a single energy transfer operation at a time, we propose an algorithm that seeks to minimize the number of mini slots that are used. First of all, an energy route for the source-consumer pair $(i,j)$ is defined as the collection of intermediate nodes to visit when transferring energy from $i$ to $j$. The algorithm proceeds as follows: 1) a route $r_{ij}$ is identified for each source $i$ and consumer $j$ (note that for the given network topology this route is {\it unique}), 2) the disjoint routes, with no power links in common, are found and are allocated to as many $(i,j)$ pairs as possible, 3) for each of these pairs $(i,j)$, the energy transfer is accomplished using route $r_{ij}$ for a number of mini slots $n_{ij}$, 4) when the transfer for a pair $(i,j)$ is complete, we check whether a new route is released (i.e., no longer used and available for subsequent transfers). If that is the case, and if this route can be used to transfer energy for any of the remaining pairs $(i^\prime, j^\prime)$ (not yet considered), this route is allocated to any of the eligible pairs $(i^\prime, j^\prime)$ for $n_{i^\prime, j^\prime}$ further mini slots. This process is repeated until all source-consumer pairs have completed their transfer.


\section{Numerical Results}
\label{sec:results}

\begin{table}[t]
	\renewcommand{\arraystretch}{1.3}
	\caption{System parameters used in the numerical results.}
	\label{tab:parameters}
	\centering
	\begin{tabular}{|l|r|}
		\hline
		Parameter 									& Value \\
		\hline
		Number of BSs, $n_s$							& $18$ \\
		Cable resistivity, $\rho$							&  $\SI{0.023}{\Omega\milli\milli}^2$/$\SI{}{\meter}$ \\
		Cable \mbox{cross-section}, $A$					& $\SI{10}{mm}^2$ \\
		Length of a power link, $\ell$						& $\SI{100}{\meter}$ \\
		Maximum energy buffer capacity, $\bmax$			& $\SI{360}{\kilo\joule}$ \\
		Upper energy threshold, $\bup$					& $0.7\bmax$ ($70\%$) \\
		Reference energy threshold, $\bref$					& $0.5\bmax$ ($50\%$) \\
		Lower energy threshold, $\blow$					& $0.1\bmax$ ($10\%$) \\
		Mini slot duration								& $\SI{60}{s}$ \\
		Maximum transmission energy capacity, $e_{max}$		& $\SI{90}{\kilo\joule}$/mini-slot \\
		MPC optimization horizon $M$						& $\SI{24}{h}$ \\
		MPC weight parameter $\alpha$					& $0.5$ \\
		Energy allocation weight parameter $\beta$ 			& $0.5$\\
		\hline
	\end{tabular}
\end{table}

In this section, the following schemes are compared: i) no energy exchange \textbf{(NOEE)}, i.e., the offline \acp{BS} only have to rely on the locally harvested energy, ii) convex solution \textbf{(CONV)}: this is the energy allocation scheme of~\cite{Gambin2017}, where the energy allocation is solely computed based on the system configuration in the current time slot. This approach is {\it myopic}, as no knowledge into the future behavior of the system is exploited. iii) Hungarian solution \textbf{(HUNG)}: the energy allocation is found through the Hungarian method of Section~\ref{sub:convex_optimization}; this is also a myopic approach. iv) Convex solution with model predictive control \textbf{(GPs+MPC+CONV)}: this is the combined optimization approach of Sections~\ref{sub:prediction},~\ref{sub:model_predictive_control} and~\ref{sub:convex_optimization}, and v) Hungarian solution with model predictive control \textbf{(GPs+MPC+HUNG)}. ii) and iii) carry out energy allocation and routing only considering the current time slot, while iv) and v) also take into account the future system evolution, exploiting pattern learning and \mbox{multi-step} ahead adaptive control.

Before discussing the numerical results, some considerations are in order. All the algorithms purchase some energy from the power grid, although the way in which they use it differs. With NOEE, the energy purchased in solely used to power the base stations that are ongrid, whereas those being offgrid have to uniquely rely on the harvested energy. Convex and Hungarian solutions allow some energy redistribution among the base stations. With these schemes, an energy rich \ac{BS} may transfer energy to other \acp{BS} whose energy buffer is depleted. Note that an energy rich base station may belong to either the ongrid set or to the offgrid one. The latter case occurs when, for instance, a \ac{BS} experiences no traffic during the day and all the energy it harvests is stored locally. In this case, this \ac{BS} is likely to be ``energy rich'', and energy transfer schemes consider it as an energy source for other \acp{BS}. Looking at the whole \ac{BS} network as a close system, it can gathers energy in two ways: i) harvesting it from the environment and ii) purchasing it from the power grid. The harvested energy is basically free of charge and shall be utilized to the best extent: energy transfer among \acp{BS} makes this possible. The energy bought by the online \acp{BS} is costly and shall also be utilized as efficiently as possible. Below, we shall evaluate both aspects.

\begin{figure}[t]	
	\centering
	\includegraphics[width=\columnwidth]{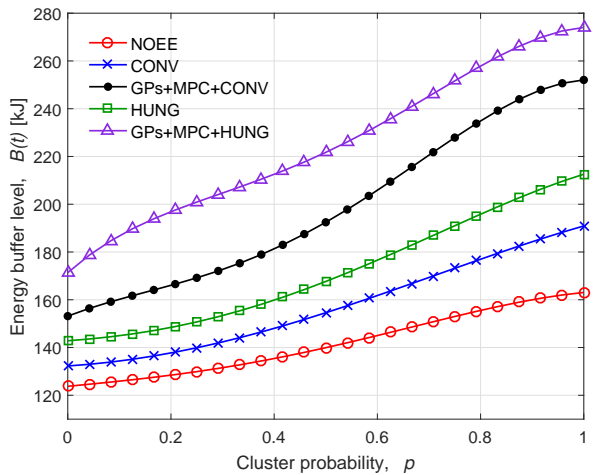}
	\caption{Energy buffer level {\it vs} cluster probability $p$.}
	\label{fig:clusterCompBattery}	
\end{figure}

\begin{figure}[t]	
	\centering
	\includegraphics[width=\columnwidth]{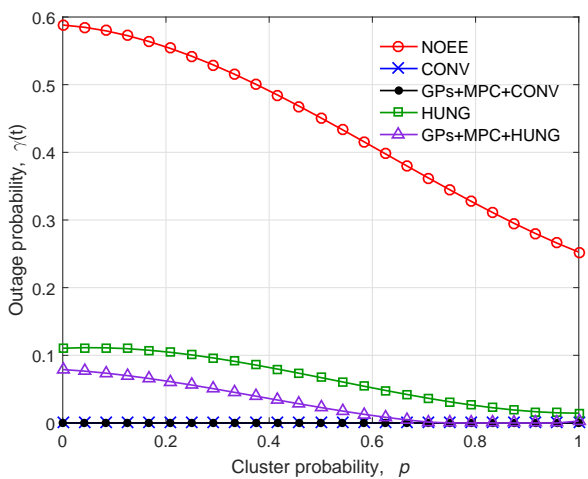}
	\caption{Outage probability $\po(t)$ {\it vs} cluster probability $p$.}
	\label{fig:clusterCompOutage}	
\end{figure}

\begin{figure}[t]	
	\centering
	\includegraphics[width=\columnwidth]{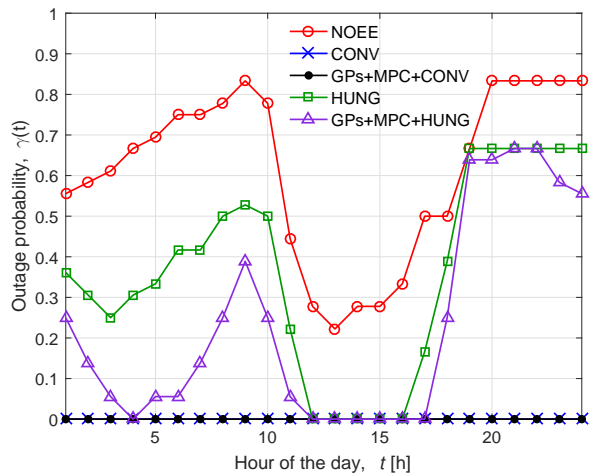}
	\caption{Outage probability $\po(t)$ over a day.}
	\label{fig:outage_prob}	
\end{figure}

\begin{figure}[t]	
	\centering
	\includegraphics[width=\columnwidth]{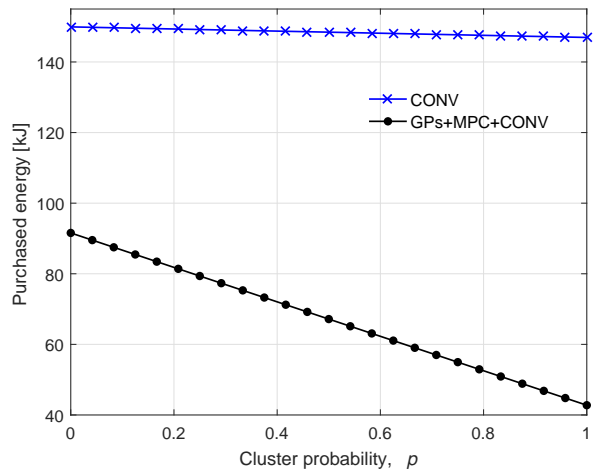}
	\caption{Purchased energy {\it vs} cluster probability $p$.}
	\label{fig:purchasedEnergy}	
\end{figure}

\begin{figure}[t]	
	\centering
	\includegraphics[width=\columnwidth]{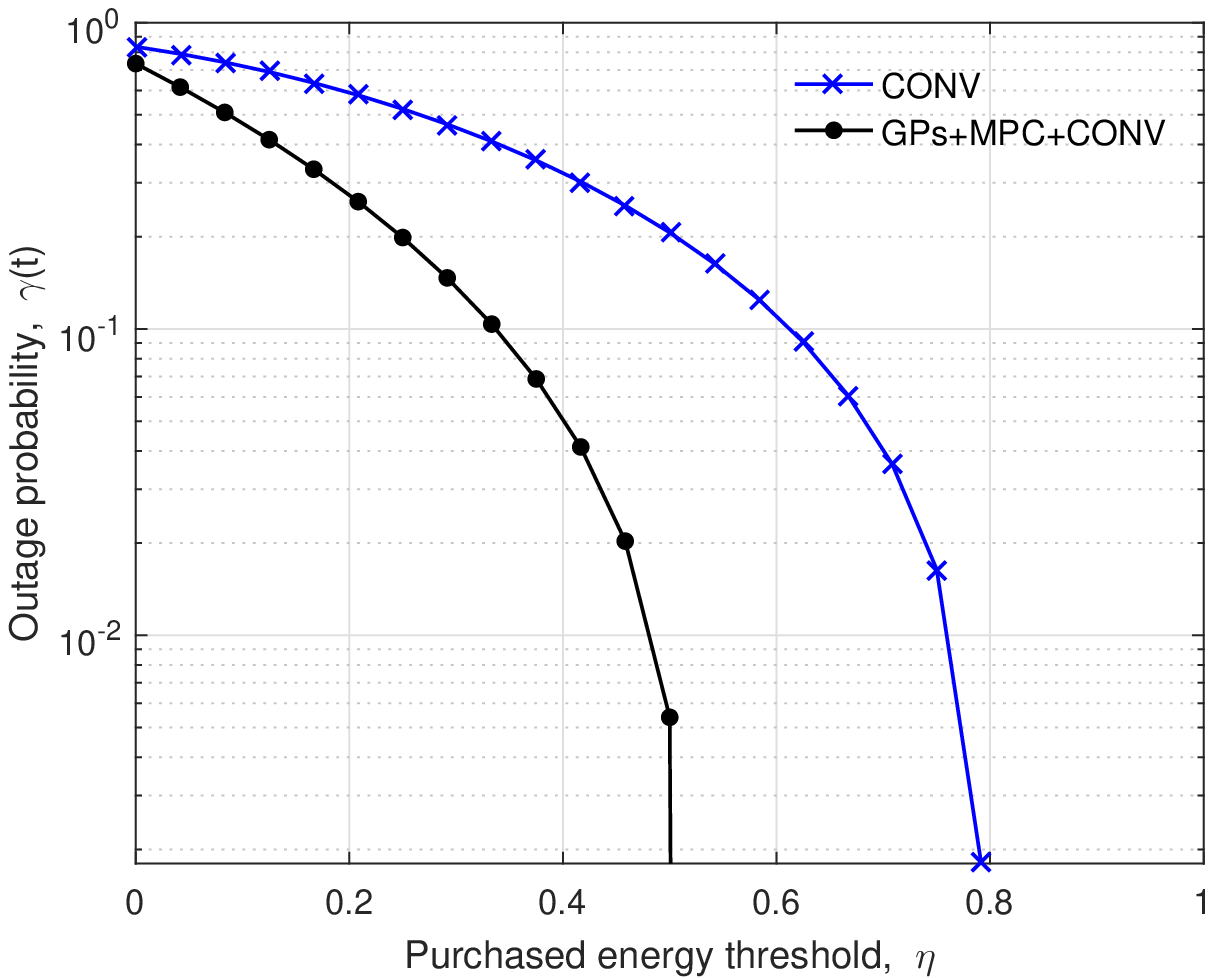}
	\caption{Outage probability $\po$ {\it vs} purchased energy threshold.}
	\label{fig:purchasedThOutage}	
\end{figure}

For the following results, we consider the scenario of Section~\ref{sec:systemModel}. For the \acp{EB}, we set $\bmax=\SI{360}{\kilo\joule}$, which corresponds to a battery capacity of $\SI{100}{\watt\hour}$ (e.g., a small size \mbox{Li-Ion} battery). The slot time is set to one hour, solar EH traces were obtained using~\cite{miozzo2014solarstat} for the city of Chicago, and the remaining simulation parameters are listed in Table~\ref{tab:parameters}.

In Fig.~\ref{fig:clusterCompBattery}, we show the average \ac{BS} energy buffer level over different traffic load configurations, considering $6$ ongrid \acp{BS} and $n_s = 18$. For the load assignment, each \ac{BS} independently picks one of the two traffic clusters in Section~\ref{sub:traffic_load}: cluster $2$ (low traffic load) is picked  with probability $p$ and cluster $1$ (high load) is picked with probability $1-p$.
As expected, the average energy buffer level when $p = 0$ is lower than that with $p = 1$, as the traffic load in cluster $1$ is higher. Regarding the approaches, the highest difference in the energy buffer levels is found between NOEE and \mbox{GPs+MPC+HUNG}, with an increment of around $60$\% (on average) when \ac{MPC} is adopted. Moreover, the Hungarian methods outperform convex solutions because, with their assignment policy, any consumer is matched to a single source and this reduces the amount of energy that is distributed, leaving more energy in the energy rich buffers. As we show shortly, this behavior is not really desirable as, e.g., it leads to higher outage probabilities.

As a proxy to the network \ac{QoS}, the outage probability at time $t$, $\po(t)$, is here defined as the ratio between the number of \acp{BS} whose energy buffer level is completely depleted, and the total number of \acp{BS} in the system $n_s$. The outage probability $\po(t)$ as a function of the traffic load is plotted in Fig.~\ref{fig:clusterCompOutage} for $|\mathcal{N}_{on}| = 6$, $n_s = 18$. For all schemes, $\po(t)$ is an increasing function of the load. The probability that a \ac{BS} runs out of service due to energy scarcity is higher when energy cannot be transferred among \acp{BS} (NOEE) and is in general very high across the whole day for \mbox{HUNG-based} solutions. However, applying \ac{MPC} to the Hungarian method leads to a reduction in the average outage probability of about $54$\%. Moreover, from Fig.~\ref{fig:outage_prob} we see that with the Hungarian method $\po(t)$ increases when the amount of energy harvested is very little (i.e., nighttime). The problem is that the Hungarian allocation technique returns a matching of \mbox{source-consumer} pairs, where each source is allocated to a single consumer and, in turn, some of the \acp{BS} may not be allocated in some time slots (due to the imbalance between number of sources and number of consumers). This leads to high outage probabilities for the considered scenario. \mbox{CONV-based} techniques are more flexible in this respect, as they allow energy transfer from multiple \acp{BS} and in different amounts. This translates into a zero outage probability in both cases, with and without \ac{MPC}. 

From the previous graphs, one may conclude that CONV and \mbox{GPs+MPC+CONV} (foresighted optimization) provide the same benefits, being both capable of lowering the outage probability down to zero. However, looking at additional metrics reveals that the two approaches show important differences. For example, in Fig.~\ref{fig:purchasedEnergy} we compare these solutions in terms of amount of energy that ongrid \acp{BS} purchase from the power grid. A big gap can be  observed between the two schemes, proving that the application of pattern learning and \ac{MPC} is indeed highly beneficial, leading to a reduction of more that $55$\% in the amount of energy purchased from the power grid.

Along these lines, we perform another set of simulations by putting a cap on the maximum amount of energy that can be bought during a full day by the ongrid \acp{BS}. Specifically, we define a {\it purchased energy threshold} $\eta$ as the ratio between the amount of energy that each ongrid BS is allowed to purchase and the total amount of energy it would require to serve a fully loaded scenario across an entire day. A plot of $\po(t)$ against threshold $\eta$ is shown in Fig.~\ref{fig:purchasedThOutage}. From this graph, we see that adaptive control (\mbox{GPs+MPC+CONV}) leads to a much smaller outage probability than CONV. Moreover, as $\eta$ increases beyond $0.5$ the outage probability drops to zero, which is a big improvement with respect to CONV, for which $\gamma$ is about $10$\%. Similar results are obtained for \mbox{GPs+MPC+HUNG} when compared with HUNG, although in this case the gain is slightly smaller. These results are not shown in the interest of space.

\section{Conclusions}
\label{sec:conclusions}

In this paper, we have considered future small cell deployments where energy harvesting and packet power networks are combined to provide energy \mbox{self-sustainability} through the use of \mbox{own-generated} energy and carefully planned power transfers among network elements. This amounts to a combined learning and optimization problem (resource scheduling), where learning is carried out on energy arrival (harvested ambient energy) and traffic load traces and this knowledge is then exploited, at runtime, for the computation of optimal energy transfer policies (among the distributed energy buffers). This (foresighted) optimization is performed combining model predictive control and convex optimization techniques. Numerical results reveal great advantages over the case where energy transfer schedules are optimized disregarding future energy and load forecasts: the amount of energy purchased from the power grid is reduced of more than $50$\% and the outage probability is reduced to zero in nearly all scenarios.  


\bibliographystyle{IEEEtran}
\bibliography{biblio}

\end{document}